\documentclass[11pt,a4paper]{article}
\pdfoutput=1
\usepackage{jcappub}
\usepackage{ifthen}
\usepackage{epsfig}
\usepackage{booktabs} 
\usepackage{natbib}
\usepackage{bbold}

\newcommand{\erf}{\mathop{\mathrm{erf}}}
\newcommand{\beqra}{\begin{flalign}}
\newcommand{\eeqra}{\end{flalign}}
\newcommand{\beq}{\begin{equation}}
\newcommand{\eeq}{\end{equation}}
\usepackage{ccicons} %new
\usepackage{listings}%new
\usepackage{color}   %new
\usepackage{braket}  %new
\usepackage{hyperref}%new
\usepackage{dsfont}  %new
\usepackage{leftidx} %new
\usepackage{bm}
\allowdisplaybreaks

\title{Dark matter directional detection in non-relativistic effective theories}
\author[a]{Riccardo Catena}
\affiliation[a]{Institut f\"ur Theoretische Physik, Friedrich-Hund-Platz 1, 37077 G\"ottingen, Germany}
\emailAdd{riccardo.catena@theorie.physik.uni-goettingen.de}

\abstract{We extend the formalism of dark matter directional detection to arbitrary one-body dark matter-nucleon interactions.~The new theoretical framework generalizes the one currently used, which is based on 2 types of dark matter-nucleon interaction only. It includes 14 dark matter-nucleon interaction operators, 8 isotope-dependent nuclear response functions, and the Radon transform of the first 2 moments of the dark matter velocity distribution. We calculate the recoil energy spectra at dark matter directional detectors made of CF$_4$, CS$_2$ and $^{3}$He for the 14 dark matter-nucleon interactions, using nuclear response functions recently obtained through numerical nuclear structure calculations.~We highlight the new features of the proposed theoretical framework, and present our results for a spherical dark matter halo and for a stream of dark matter particles.~This study lays the foundations for model independent analyses of dark matter directional detection experiments.}

\begin{document}
\maketitle

\section{Introduction}
The Earth's motion with respect to the galactic rest frame is expected to produce a flux of dark matter particles across the surface of the planet~\cite{Rubin:1970}. Fixed target experiments can in principle detect these particles, if they scatter and deposit energy in low-background detectors~\cite{Goodman:1984dc,Drukier:1986tm,Freese:1987wu}.

The angular distribution of nuclear recoil events from dark matter-nucleus scattering 
at fixed target experiments is not isotropic, as the Earth's motion selects a preferred direction in the sphere of recoil directions~\cite{Spergel:1987kx}. Recoil events are mainly expected in the direction opposite to the observer's motion, or in a ring around it for large dark matter particle mass to recoil energy ratios~\cite{Bozorgnia:2011vc}. 

Dark matter directional detectors are designed to measure anisotropies in the distribution of nuclear recoil events at fixed target experiments. Specifically, their task is to measure the recoil momentum vector of nuclei scattered off by dark matter particles from the local galactic population~\cite{Ahlen:2009ev}. The sense of the recoil momentum vector is also called the ``head-tail'' of the recoil track. 

Dark matter directional detection experiments currently in a research and development stage are DRIFT~\cite{Battat:2014van,Daw:2011wq}, MIMAC~\cite{Riffard:2013psa,Santos:2013hpa}, DMTPC~\cite{Battat:2012,Monroe:2012qma}, NEWAGE~\cite{Miuchi:2012rma,Miuchi:2010hn} and D3~\cite{Vahsen:2011qx}. They adopt diffuse gas detectors and time projection chambers to reconstruct the nuclear recoil tracks. Alternative approaches include the use of nuclear emulsions~\cite{Naka:2011sf}, dark matter-electron scattering in crystals~\cite{Essig:2011nj} and DNA detectors~\cite{Drukier:2012hj}.
In the near future, directional detection experiments should reach the sensitivity to detect standard weakly interacting massive particles (WIMPs), as those predicted by SUSY~\cite{Jungman:1995df,Catena:2013pka,Catena:2009tm,Catena:2004ba} or extra-dimensional theories~\cite{Servant:2002aq,Hooper:2007qk,Buchmuller:2008cf,Buchmuller:2009er}.

The recoil energy spectrum expected at dark matter directional detectors has so far been computed for two types of dark matter-nucleon interaction only, e.g.~\cite{Gondolo:2002np,Morgan:2004ys,Green:2006cb,Alenazi:2007sy,Green:2007at,Green:2010zm,Billard:2011zj,Grothaus:2014hja,Kavanagh:2015aqa,Laha:2015yoa}. One is the familiar spin-independent interaction. The other one is the well-known spin-dependent dark matter-nucleon interaction. At the quantum mechanical level, both interactions are independent of the momentum transfer operator, and of the dark matter-nucleon relative velocity operator. 

Though assuming momentum and velocity independent dark matter-nucleon interactions is a reasonable first approximation, it only provides a limited, and to some extent biased description of the actual complexity of the dark matter-nucleus scattering~\cite{Catena:2014hla}. The most general non-relativistic effective theory for one-body dark matter-nucleon interactions predicts 14 Galilean invariant dark matter-nucleon interaction operators, each with an isoscalar and an isovector component~\cite{Fitzpatrick:2012ix,Anand:2013yka}. On purely observational grounds, there is no reason to neglect any of the 14 interaction operators, which motivates the exploration of more general approaches.

In this work we extend the formalism of dark matter directional detection to arbitrary one-body dark matter-nucleon interactions. The ultimate goal of this study is to set the bases for fully model-independent analyses of future dark matter directional detection experiments. Non-relativistic dark matter-nucleon interaction operators were previously explored in the context of dark matter direct detection in~\cite{Chang:2009yt,Fan:2010gt,Fornengo:2011sz,Fitzpatrick:2012ix,Fitzpatrick:2012ib,Menendez:2012tm,Cirigliano:2012pq,Anand:2013yka,DelNobile:2013sia,Klos:2013rwa,Peter:2013aha,Hill:2013hoa,Catena:2014uqa,Catena:2014hla,Catena:2014epa,Gluscevic:2014vga,Panci:2014gga,Vietze:2014vsa,Barello:2014uda,Schneck:2015eqa,Catena:2015uua,Scopel:2015baa}, and studying the dark matter capture by the Sun in~\cite{Guo:2013ypa,Liang:2013dsa,Blumenthal:2014cwa,Vincent:2013lua,Lopes:2014aoa,Vincent:2014jia,Catena:2015iea,Vincent:2015gqa}.

The paper is organized as follows. In Sec.~\ref{sec:theory} we introduce the non-relativistic effective theory of dark matter directional detection. The theory is carefully explored in Sec.~\ref{sec:pheno} for three benchmark detectors and two astrophysical configurations. We summarize our results in Sec.~\ref{sec:conc}, and list useful equations in the Appendixes. 

\section{Effective theory of dark matter directional detection}
\label{sec:theory}
In this section we extend the formalism of dark matter directional detection to arbitrary one-body dark matter-nucleon interactions. 

\subsection{Dark matter-nucleus scattering in effective theories}
\label{sec:EFT}
We start with a brief review of the effective theory of one-body dark matter-nucleon interactions~\cite{Fitzpatrick:2012ix}. It provides a general framework for the study of dark matter scattering from target nuclei at directional detectors.

Under the assumption of one-body dark matter-nucleon interactions, the Hamiltonian density 
\begin{equation}
\hat{\mathcal{H}}_{\rm T}= \sum_{i=1}^{A}  \sum_{\tau=0,1} \sum_{k} c_k^{\tau}\hat{\mathcal{O}}_{k}^{(i)} \, t^{\tau}_{(i)} \,
\label{eq:H_I}
\end{equation}
describes the most general short-range dark matter-nucleus interaction~\cite{Fitzpatrick:2012ix}.
In Eq.~(\ref{eq:H_I}), the sum over $i=1,\dots, A$, where $A$ is the target nucleus mass number, reflects the assumption of one-body interactions. The 14 dark matter-nucleon interaction operators $\hat{\mathcal{O}}_{k}^{(i)}$ depend on the momentum transfer operator ${\bf{\hat{q}}}$, on the relative transverse velocity operator ${\bf{\hat{v}}}^{\perp}$, and on the dark matter particle and nucleon spin operators, ${\bf{\hat{S}}}_\chi$ and ${\bf{\hat{S}}}_N$, respectively~\cite{Fitzpatrick:2012ix}. We list them in Tab.~\ref{tab:operators}. 
Introducing the $2\times2$ matrices in isospin space $t^0=\mathbb{1}$ and $t^{1}=\tau_3$, where $\tau_3$ is the third Pauli matrix, we relate the isoscalar and isovector coupling constants, $c_k^0$ and $c_k^1$ respectively, to the coupling constants for protons ($c_k^p$) and nucleons  ($c_k^n$) as follows: $c^{p}_k=(c^{0}_k+c^{1}_k)/2$, and $c^{n}_k=(c^{0}_k-c^{1}_k)/2$. Defined in this way, $c_k^0$ and $c_k^1$ have dimension mass to the power of $-2$.
The Hamiltonian density~(\ref{eq:H_I}) admits the following coordinate space representation 
\begin{eqnarray}
\hat{\mathcal{H}}_{\rm T}({\bf{r}}) = \sum_{\tau=0,1} &\Bigg\{&
\sum_{i=1}^A  \hat{l}_0^{\tau}~ \delta({\bf{r}}-{\bf{r}}_i)
 + \sum_{i=1}^A  {\bf{\hat{l}}}_5^{\tau} \cdot \vec{\sigma}_i \,\delta({\bf{r}}-{\bf{r}}_i) \nonumber\\
 &+&   \sum_{i=1}^A {\bf{\hat{l}}}_M^{\tau} \cdot \frac{1}{2 m_N} \Bigg[i \overleftarrow{\nabla}_{\bf{r}}\delta({\bf{r}}-{\bf{r}}_i) -i \delta({\bf{r}}-{\bf{r}}_i)\overrightarrow{\nabla}_{\bf{r}} \Bigg]  \nonumber \\
&+& \sum_{i=1}^A {\bf{\hat{l}}}_E^{\tau} \cdot \frac{1}{2m_N} \Bigg[ \overleftarrow{\nabla}_{\bf{r}} \times \vec{\sigma}_i \,\delta({\bf{r}}-{\bf{r}}_i) +\delta({\bf{r}}-{\bf{r}}_i)\
  \vec{\sigma}_i \times \overrightarrow{\nabla}_{\bf{r}} \Bigg] \Bigg\} t^{\tau}_{(i)} \,.\nonumber\\
\label{eq:Hx}
\end{eqnarray}
It depends on the Pauli matrices that represent the $i$th-nucleon spin operator $\vec{\sigma}_i$, on the dark matter-nucleus relative distance ${\bf{r}}$, and on the $i$th-nucleon position in the nucleus center of mass frame ${\bf{r}}_i$~\cite{Fitzpatrick:2012ix}.

\begin{table}[t]
    \centering
    \begin{tabular}{ll}
    \toprule
        $\hat{\mathcal{O}}_1 = \mathbb{1}_{\chi N}$ & $\hat{\mathcal{O}}_9 = i{\bf{\hat{S}}}_\chi\cdot\left({\bf{\hat{S}}}_N\times\frac{{\bf{\hat{q}}}}{m_N}\right)$  \\
        $\hat{\mathcal{O}}_3 = i{\bf{\hat{S}}}_N\cdot\left(\frac{{\bf{\hat{q}}}}{m_N}\times{\bf{\hat{v}}}^{\perp}\right)$ \hspace{2 cm} &   $\hat{\mathcal{O}}_{10} = i{\bf{\hat{S}}}_N\cdot\frac{{\bf{\hat{q}}}}{m_N}$   \\
        $\hat{\mathcal{O}}_4 = {\bf{\hat{S}}}_{\chi}\cdot {\bf{\hat{S}}}_{N}$ &   $\hat{\mathcal{O}}_{11} = i{\bf{\hat{S}}}_\chi\cdot\frac{{\bf{\hat{q}}}}{m_N}$   \\                                                                             
        $\hat{\mathcal{O}}_5 = i{\bf{\hat{S}}}_\chi\cdot\left(\frac{{\bf{\hat{q}}}}{m_N}\times{\bf{\hat{v}}}^{\perp}\right)$ &  $\hat{\mathcal{O}}_{12} = {\bf{\hat{S}}}_{\chi}\cdot \left({\bf{\hat{S}}}_{N} \times{\bf{\hat{v}}}^{\perp} \right)$ \\                                                                                                                 
        $\hat{\mathcal{O}}_6 = \left({\bf{\hat{S}}}_\chi\cdot\frac{{\bf{\hat{q}}}}{m_N}\right) \left({\bf{\hat{S}}}_N\cdot\frac{\hat{{\bf{q}}}}{m_N}\right)$ &  $\hat{\mathcal{O}}_{13} =i \left({\bf{\hat{S}}}_{\chi}\cdot {\bf{\hat{v}}}^{\perp}\right)\left({\bf{\hat{S}}}_{N}\cdot \frac{{\bf{\hat{q}}}}{m_N}\right)$ \\   
        $\hat{\mathcal{O}}_7 = {\bf{\hat{S}}}_{N}\cdot {\bf{\hat{v}}}^{\perp}$ &  $\hat{\mathcal{O}}_{14} = i\left({\bf{\hat{S}}}_{\chi}\cdot \frac{{\bf{\hat{q}}}}{m_N}\right)\left({\bf{\hat{S}}}_{N}\cdot {\bf{\hat{v}}}^{\perp}\right)$  \\
        $\hat{\mathcal{O}}_8 = {\bf{\hat{S}}}_{\chi}\cdot {\bf{\hat{v}}}^{\perp}$  & $\hat{\mathcal{O}}_{15} = -\left({\bf{\hat{S}}}_{\chi}\cdot \frac{{\bf{\hat{q}}}}{m_N}\right)\left[ \left({\bf{\hat{S}}}_{N}\times {\bf{\hat{v}}}^{\perp} \right) \cdot \frac{{\bf{\hat{q}}}}{m_N}\right] $ \\                                                                               
    \bottomrule
    \end{tabular}
    \caption{Non-relativistic quantum mechanical operators defining the general effective theory of one-body dark matter-nucleon interactions. All operators have the same mass dimension, and $m_N$ is the nucleon mass. Here and in the next sections we omit the nucleon index $(i)$.} 
    \label{tab:operators}
\end{table}
Eq.~(\ref{eq:Hx}) describes the dark matter coupling to the nuclear vector charge (first term in the first line), to the nuclear spin-current (second term in the first line), to the nuclear convection current (second line), and finally, to the nuclear spin-velocity current (last line). The exact nature of these couplings is encoded in the four operators $\hat{l}_0^{\tau}$, ${\bf{\hat{l}}}_5^{\tau}$, ${\bf{\hat{l}}}_M^{\tau}$, and ${\bf{\hat{l}}}_E^{\tau}$. They are defined as follows
\begin{align}
\label{eq:ls}
\hat{l}_0^\tau &= c_1^\tau + i  \left( {{\bf{\hat{q}}} \over m_N}  \times {\bf{\hat{v}}}_{T}^\perp \right) \cdot  {\bf{\hat{S}}}_\chi  ~c_5^\tau
+ {\bf{\hat{v}}}_{T}^\perp \cdot {\bf{\hat{S}}}_\chi  ~c_8^\tau + i {{\bf{\hat{q}}} \over m_N} \cdot {\bf{\hat{S}}}_\chi ~c_{11}^\tau \nonumber \\
{\bf{\hat{l}}}_5^{\tau}&= {1 \over 2} \left[ i {{\bf{\hat{q}}} \over m_N} \times {\bf{\hat{v}}}_{T}^\perp~ c_3^\tau + {\bf{\hat{S}}}_\chi ~c_4^\tau
+  {{\bf{\hat{q}}} \over m_N}~{{\bf{\hat{q}}} \over m_N} \cdot {\bf{\hat{S}}}_\chi ~c_6^\tau
+   {\bf{\hat{v}}}_{T}^\perp ~c_7^\tau + i {{\bf{\hat{q}}} \over m_N} \times {\bf{\hat{S}}}_\chi ~c_9^\tau + i {{\bf{\hat{q}}} \over m_N}~c_{10}^\tau \right. \nonumber \\
&+  \left. {\bf{\hat{v}}}_{T}^\perp \times {\bf{\hat{S}}}_\chi ~c_{12}^\tau
+i  {{\bf{\hat{q}}} \over m_N} {\bf{\hat{v}}}_{T}^\perp \cdot {\bf{\hat{S}}}_\chi ~c_{13}^\tau+i {\bf{\hat{v}}}_{T}^\perp {{\bf{\hat{q}}} \over m_N} \cdot {\bf{\hat{S}}}_\chi ~ c_{14}^\tau+{{\bf{\hat{q}}} \over\
 m_N} \times {\bf{\hat{v}}}_{T}^\perp~ {{\bf{\hat{q}}} \over m_N} \cdot {\bf{\hat{S}}}_\chi ~ c_{15}^\tau  \right]\nonumber \\
{\bf{\hat{l}}}_M^{\tau} &=   i {{\bf{\hat{q}}} \over m_N}  \times {\bf{\hat{S}}}_\chi ~c_5^\tau - {\bf{\hat{S}}}_\chi ~c_8^\tau \nonumber \\
{\bf{\hat{l}}}_E^{\tau} &= {1 \over 2} \left[  {{\bf{\hat{q}}} \over m_N} ~ c_3^\tau +i {\bf{\hat{S}}}_\chi~c_{12}^\tau - {{\bf{\hat{q}}} \over  m_N} \times{\bf{\hat{S}}}_\chi  ~c_{13}^\tau-i 
{{\bf{\hat{q}}} \over  m_N} {{\bf{\hat{q}}} \over m_N} \cdot {\bf{\hat{S}}}_\chi  ~c_{15}^\tau \right] \,. \nonumber\\
\end{align}
The four operators $\hat{l}_0^{\tau}$, ${\bf{\hat{l}}}_5^{\tau}$, ${\bf{\hat{l}}}_M^{\tau}$, and ${\bf{\hat{l}}}_E^{\tau}$ depend on the dark matter particle spin operator ${\bf{\hat{S}}}_\chi$, on the momentum transfer operator ${\bf{\hat{q}}}$, and on the operator ${\bf{\hat{v}}}^{\perp}_T={\bf{\hat{v}}}^{\perp}-{\bf{\hat{v}}}^{\perp}_N$, where $2 m_N {\bf{\hat{v}}}^{\perp}_N=i \overleftarrow{\nabla}_{{\bf{r}}} \,\delta({\bf{r}}-{\bf{r}}_i) - i\delta({\bf{r}}-{\bf{r}}_i) \overrightarrow{\nabla}_{{\bf{r}}}$. Analogous coordinate space representations for ${\bf{\hat{q}}}$ and ${\bf{\hat{v}}}^{\perp}_T$ can be found in~\cite{Catena:2015uha}.

The Hamiltonian density in Eq.~(\ref{eq:Hx}) leads to the following transition probability for dark matter-nucleus scattering
\begin{align}
\langle |\mathcal{M}_{T}|^2\rangle_{\rm spins} =  \frac{4\pi}{2J+1}\sum_{\tau,\tau'} &\bigg[ \sum_{k=M,\Sigma',\Sigma''} R^{\tau\tau'}_k\left(v_T^{\perp 2}, {q^2 \over m_N^2} \right) W_k^{\tau\tau'}(y) \nonumber\\
&+{q^{2} \over m_N^2} \sum_{k=\Phi'', \Phi'' M, \tilde{\Phi}', \Delta, \Delta \Sigma'} R^{\tau\tau'}_k\left(v_T^{\perp 2}, {q^2 \over m_N^2}\right) W_k^{\tau\tau'}(y) \bigg] \,, \nonumber\\
\label{eq:M}
\end{align}
where $M_{T}$ is the transition amplitude, $J$ is the nuclear spin, and the angle brackets denote a sum (average) over the final (initial) spin-configurations. The index $k$ in Eq.~(\ref{eq:M}) extends over the 8 nuclear response functions $W_k^{\tau\tau'}(y)$ defined below in Sec.~\ref{sec:nuc} together with the new kinematical variable $y$.

We refer to the 8 functions $R^{\tau\tau'}_k$ in Eq.~(\ref{eq:M}) as dark matter response functions. They are quadratic in matrix elements of the operators in Eq.~(\ref{eq:ls}), and depend on $q^2/m_N^2$ and on $v_T^{\perp 2}=v^2-q^2/(4\mu_T^2)$, where $v$ is the dark matter-nucleus relative velocity, $m_N$ is the nucleon mass, and $\mu_T$ the reduced dark matter-nucleus mass. We list the functions $R^{\tau\tau'}_k$ in Appendix~\ref{sec:appDM}.

The differential cross-section for dark matter scattering from nuclei of type $T$ and mass $m_T$ can finally be written as follows
\begin{equation}
\frac{{\rm d}\sigma_T(q^2,v^2)}{{\rm d}q^2} = \frac{1}{4\pi v^2} \, \langle |\mathcal{M}_{T}|^2\rangle_{\rm spins} \,,
\label{eq:sigma}
\end{equation}
which for arbitrary interactions is a function of the momentum transfer, and of the dark matter-nucleus relative velocity.

\subsection{Nuclear response functions and target materials}
\label{sec:nuc}
The 8 nuclear response functions in Eq.~(\ref{eq:M}) arise from a multipole expansion of the nuclear charge and currents in Eq.~(\ref{eq:Hx}), and are defined as follows
\begin{equation}
W_{\alpha\beta}^{\tau \tau^\prime}(y)= \sum_{L}  \langle J,T,M_T ||~ \alpha_{L;\tau} (q)~ || J,T,M_T \rangle \langle J,T,M_T ||~ \beta_{L;\tau^\prime} (q)~ || J,T,M_T \rangle \,.
\label{eq:W}
\end{equation}
The multiple expansion index $L$ must be less then $2J$.
In Eq.~(\ref{eq:W}), we label the state $|J,T,M_T \rangle$ using the nuclear spin $J$, the nuclear isospin $T$, and the isospin magnetic quantum number $M_T$.
The functions $W_{\alpha\beta}^{\tau \tau^\prime}$ are therefore quadratic in nuclear matrix elements reduced in the nuclear spin magnetic quantum number. Assuming the harmonic oscillator basis for single-nucleon states, the nuclear response functions $W_{\alpha\beta}^{\tau \tau^\prime}$ depend on $y=(bq/2)^2$ only, where
\begin{align}
b=\sqrt{41.467/(45 A^{-1/3}-25A^{-2/3})}~{\rm fm}\,.
\end{align} 
The nuclear response operators $\alpha_{LM;\tau}$ and $\beta_{LM;\tau}$ in Eq.~(\ref{eq:W}), can each be one of the following operators
\begin{eqnarray}
M_{LM;\tau}(q) &=& \sum_{i=1}^{A} M_{LM}(q {\bf{r}}_i) t^{\tau}_{(i)}\nonumber\\
\Sigma'_{LM;\tau}(q) &=& -i \sum_{i=1}^{A} \left[ \frac{1}{q} \overrightarrow{\nabla}_{{\bf{r}}_i} \times {\bf{M}}_{LL}^{M}(q {\bf{r}}_i)  \right] \cdot \vec{\sigma}_i \, t^{\tau}_{(i)}\nonumber\\
\Sigma''_{LM;\tau}(q) &=&\sum_{i=1}^{A} \left[ \frac{1}{q} \overrightarrow{\nabla}_{{\bf{r}}_i} M_{LM}(q {\bf{r}}_i)  \right] \cdot \vec{\sigma}_i \, t^{\tau}_{(i)}\nonumber\\
\Delta_{LM;\tau}(q) &=&\sum_{i=1}^{A}  {\bf{M}}_{LL}^{M}(q {\bf{r}}_i) \cdot \frac{1}{q}\overrightarrow{\nabla}_{{\bf{r}}_i} t^{\tau}_{(i)} \nonumber\\
\tilde{\Phi}^{\prime}_{LM;\tau}(q) &=& \sum_{i=1}^A \left[ \left( {1 \over q} \overrightarrow{\nabla}_{{\bf{r}}_i} \times {\bf{M}}_{LL}^M(q {\bf{r}}_i) \right) \cdot \left(\vec{\sigma}_i \, \times {1 \over q} \overrightarrow{\nabla}_{{\bf{r}}_i} \right) + {1 \over 2} {\bf{M}}_{LL}^M(q {\bf{r}}_i) \cdot \vec{\sigma}_i \, \right]t^\tau_{(i)} \nonumber \\
\Phi^{\prime \prime}_{LM;\tau}(q) &=& i  \sum_{i=1}^A\left( {1 \over q} \overrightarrow{\nabla}_{{\bf{r}}_i}  M_{LM}(q {\bf{r}}_i) \right) \cdot \left(\vec{\sigma}_i \, \times \
{1 \over q} \overrightarrow{\nabla}_{{\bf{r}}_i}  \right)t^\tau_{(i)} \,,
\label{eq:multipole}
\end{eqnarray}
where $M_{LM}(q {\bf{r}}_i)=j_{L}(q r_i)Y_{LM}(\Omega_{{\bf{r}}_i})$, and ${\bf{M}}_{LL}^{M}(q {\bf{r}}_i)=j_{L}(q r_i){\bf Y}^M_{LL1}(\Omega_{{\bf{r}}_i})$. The vector spherical harmonics, ${\bf Y}^M_{LL1}(\Omega_{{\bf{r}}_i})$, are defined as follows 
\begin{equation}
{\bf Y}^M_{LL'1}(\Omega_{{\bf{r}}_i}) = \sum_{m\lambda} \langle L'm1\lambda|L'1LM \rangle
Y_{L'm}(\Omega_{{\bf{r}}_i}) \, {\bf e}_\lambda \,,
\end{equation}
where $ \langle L'm1\lambda|L'1LM \rangle$ are Clebsch-Gordan coefficients, and ${\bf e}_\lambda$ is a spherical unit vector basis. In Eq.~(\ref{eq:M}), the index $k$ extends over pairs $\alpha\beta$ of nuclear response operators, e.g. $k= \Delta \Sigma'$. We however use the notation $W_{\alpha}^{\tau \tau^\prime}(y)\equiv W_{\alpha\beta}^{\tau \tau^\prime}(y)$ for $\alpha=\beta$. 

Dark matter directional detection experiments currently in a research and development stage mainly exploit target materials made of CF$_4$, CS$_2$ and $^3$He, or of mixtures of them. For instance, experiments like DMTPC and NEWAGE adopt detectors composed of CF$_4$, whereas DRIFT and MIMAC use, respectively, CS$_2$ (or a CS$_2$-CF$_4$ mixture) and $^3$He (or CF$_4$) as target materials.

For the elements $^{3}$He, $^{12}$C, and $^{32}$S, we use the nuclear response functions derived in~\cite{Catena:2015uha} through numerical nuclear structure calculations. For $^{19}$F we adopt the nuclear response functions found in~\cite{Fitzpatrick:2012ix}. For reference, the nuclear response functions relevant for dark matter directional detection are listed in Appendix~\ref{sec:appNuc}.

\subsection{Recoil energy spectra}
\label{sec:Rate}
For a given dark matter-nucleon interaction operator in Tab.~\ref{tab:operators}, and a given target material, the double differential nuclear recoil energy spectrum per unit time and per unit detector mass expected at a directional detection experiment is given by
\begin{eqnarray}
\frac{{\rm d}^2\mathcal{R}}{{\rm d}E_{R}\,{\rm d}\Omega} &\equiv&  \sum_{T}\frac{{\rm d}^2\mathcal{R}_{T}}{{\rm d}E_{R}\,{\rm d}\Omega} \nonumber\\
&=&  \sum_{T} \frac{\xi_T}{(2\pi)^2} \frac{\rho_{\chi}}{m_\chi}  
 \int \delta({\bf v}\cdot {\bf w}-w_T)\, f({\bf v} + {\bf v_e}(t)) \, \langle |\mathcal{M}_{T}|^2\rangle_{\rm spins} \, {\rm d}^3{\bf v} \,,
\label{rate_theory}
\end{eqnarray}
where $E_R=q^2/(2m_T)$ is the nuclear recoil energy, ${\bf w}$ the nuclear recoil direction ($|{\bf w}|^2=1$), $m_\chi$ the dark matter particle mass, $\xi_T$ the mass fraction of the nucleus $T$ in the target material, and $\rho_\chi$ the local dark matter density. In Eq.~(\ref{rate_theory}), $w_T=q/2\mu_T$ is the isotope-dependent minimum velocity required to transfer a momentum $q$ from the target nucleus to the dark matter particle, and ${\bf v_e}(t)$ is the time-dependent Earth's velocity in the galactic rest frame. The angle $\theta$ is measured with respect to the reference direction ${\bf v_e}(t)$. 
Here we assume azimuthal symmetry of ${\rm d}^2\mathcal{R}/{\rm d}E_{R}\,{\rm d}\Omega$ around the direction ${\bf v_e}(t)$, i.e. ${\rm d}\Omega=2\pi {\rm d}\hspace{-0.5mm}\cos\theta$.
The integral in Eq.~(\ref{rate_theory}) corresponds to the Radon transform $\hat{f}_M$ of the local dark matter velocity distribution $f$ in the galactic rest frame boosted to the detector frame times the transition probability (\ref{eq:M}):
\begin{equation}
\hat{f}_M(w_T,{\bf w})= \int \delta({\bf v}\cdot {\bf w}-w_T)\, f({\bf v} + {\bf v_e}(t)) \, \langle |\mathcal{M}_{T}|^2\rangle_{\rm spins} \, {\rm d}^3{\bf v} \,.
\end{equation}
Our assumptions regarding $f({\bf v} + {\bf v_e}(t))$ and ${\bf v_e}(t)$ will be discussed in the next subsection, where we also provide analytic expressions for the Radon transform $\hat{f}_M$, given the transition probability (\ref{eq:M}).

In the figures of Sec.~\ref{sec:pheno} we plot the differential number of recoil events around $\cos\theta$ per unit time and per unit detector mass:
\begin{equation}
\frac{{\rm d}\mathcal{R}}{{\rm d}\hspace{-0.5mm}\cos\theta} = 2\pi\int_{E_R>E_{\rm th}} \,\frac{{\rm d}^2\mathcal{R}}{{\rm d}E_{R}\,{\rm d}\Omega} {\rm d}E_R \,,
\label{eq:dR}
\end{equation}
where $E_{\rm th}$ is the detector energy threshold.

\subsection{Radon transforms}
\label{sec:Radon}
For $f({\bf v} + {\bf v_e}(t))$ we assume a normalized Maxwell-Boltzmann distribution truncated at the escape velocity $v_{\rm esc}$:
\begin{equation}
f({\bf v} + {\bf v_e}(t))= \frac{1}{N_{\rm esc}(2\pi \sigma_v^2)^{3/2}}\,\exp\left(-\frac{|{\bf v}+{\bf v}_{e}|^2}{2 \sigma_v^2}\right)\, \Theta\left(v_{\rm esc}-|{\bf v}+{\bf v}_{e}|\right)\,,
\label{eq:MB}
\end{equation}
where the normalization constant $N_{\rm esc}$ is given by
\begin{equation}
N_{\rm esc}= \erf\left(\frac{v_{\rm esc}}{\sqrt{2} \sigma_v}\right) - \sqrt{\frac{2}{\pi}} \frac{v_{\rm esc}}{\sigma_v} \exp\left(-\frac{v_{\rm esc}^2}{2\sigma_v^2} \right) \,,
\label{eq:Nesc}
\end{equation}
and $\sigma_v$ is the velocity dispersion. In Eqs.~(\ref{eq:MB}) and (\ref{eq:Nesc}), $v_{\rm esc}$ and ${\bf v_e}(t)$ are measured in the galactic rest frame, whereas ${\bf v}$ is the dark matter-nucleus relative velocity in the detector frame. 

Eq.~(\ref{eq:MB}) is a standard assumption for the local dark matter velocity distribution. Though a first approximation only, it allows to analytically compute $\hat{f}_M(w_T,{\bf w})$ for a given transition probability $\langle |\mathcal{M}_{T}|^2\rangle_{\rm spins}$. We leave the exploration of  velocity distributions self-consistently generated by Eddington's inversion method~\cite{Catena:2011kv}, or by its anisotropic extensions~\cite{Bozorgnia:2013pua,Fornasa:2013iaa}, for future work.

Inspection of Eq.~(\ref{eq:R}) shows that for arbitrary one-body dark matter-nucleon interactions, two independent Radon transforms appear in Eqs.~(\ref{rate_theory}) and (\ref{eq:dR}). One, $\hat{f}_M^{(0)}$, for $\langle |\mathcal{M}_{T}|^2\rangle_{\rm spins}\propto v^0$, the other one , $\hat{f}_M^{(2)}$, for $\langle |\mathcal{M}_{T}|^2\rangle_{\rm spins}\propto v^2$. 

Within the approximation~(\ref{eq:MB}), we can analytically calculate both $\hat{f}_M^{(0)}$ and $\hat{f}_M^{(2)}$. We find 
\begin{eqnarray}
\label{eq:fM0}
\hat{f}_M^{(0)}(w_T,{\bf w})&\equiv& \int \delta({\bf v}\cdot {\bf w}-w_T)\, f({\bf v} + {\bf v_e}(t)) \, {\rm d}^3{\bf v} \, \nonumber\\
&=&\frac{1}{N_{\rm esc}(2\pi \sigma_v^2)^{1/2}}\,\left\{ \exp\left[-\frac{\left(w_T+|{\bf v}_e(t)|\cos\theta\right)^2}{2\sigma_v^2}\right] - \exp\left(-\frac{v_{\rm esc}^2}{2\sigma_v^2}\right) \right\}\,,\\
&&\nonumber\\
&&\nonumber\\
\label{eq:fM2}
\hat{f}_M^{(2)}(w_T,{\bf w})&\equiv& \int \delta({\bf v}\cdot {\bf w}-w_T)\, f({\bf v} + {\bf v_e}(t)) \, v^2 \, {\rm d}^3{\bf v} \, \nonumber\\
\nonumber\\
&=& \frac{1}{N_{\rm esc}(2\pi \sigma_v^2)^{1/2}}\,\Bigg\{ \exp\left[-\frac{\left(w_T+|{\bf v}_e(t)|\cos\theta\right)^2}{2\sigma_v^2}\right] 
\left( w_T^2+|{\bf v}_e(t)|^2\sin^2\theta + 2\sigma_v^2 \right) \nonumber\\
&-& \exp\left(-\frac{v_{\rm esc}^2}{2\sigma_v^2}\right) \Bigg[ v_{\rm esc}^2 + 2 \sigma_v^2 + |{\bf v}_e(t)|^2 -2|{\bf v}_e(t)|\cos\theta\left(w_T+|{\bf v}_e(t)|\cos\theta\right) \Bigg] \Bigg\} \,.
\nonumber\\
\end{eqnarray}
for $\left(w_T+|{\bf v}_e(t)|\cos\theta\right)<v_{\rm esc}$, and zero otherwise. 
We have verified the validity of Eqs.~(\ref{eq:fM0}) and~(\ref{eq:fM2}) using the Radon transform inversion theorem, which under minimal regularity conditions relates a function $g$ to its Radon transform $\hat{g}$ as follows
\begin{equation}
g({\bf{v}}) = -\frac{1}{8\pi^2} \int \hat{g}^{\prime\prime}({\bf v}\cdot {\bf w},{\bf w}) \, {\rm d}\Omega_{w} \,.
\label{eq:invfM}
\end{equation}
In Eq.~(\ref{eq:invfM}), $\hat{g}^{\prime\prime}(x,{\bf w})\equiv\partial^2\hat{g}(x,{\bf w})/\partial x^2$, $x={\bf v}\cdot {\bf w}$, and the integral is over all recoil directions ${\bf w}$. Using Eq.~(\ref{eq:invfM}), we were able to re-derive $f({\bf v} + {\bf v_e}(t))$ and $v^2 f({\bf v} + {\bf v_e}(t))$ from Eq.~(\ref{eq:fM0}) for $\hat{f}_M^{(0)}(w_T,{\bf w})$ and Eq.~(\ref{eq:fM2}) for $\hat{f}_M^{(2)}(w_T,{\bf w})$, respectively. Our Eqs.~(\ref{eq:fM0}) and~(\ref{eq:fM2}) allow to evaluate Eqs.~(\ref{rate_theory}) and (\ref{eq:dR}) for all one-body dark matter-nucleon interaction operators in Tab.~\ref{tab:operators}.

\begin{figure}[t]
\begin{center}
\begin{minipage}[t]{0.49\linewidth}
\centering
\includegraphics[width=\textwidth]{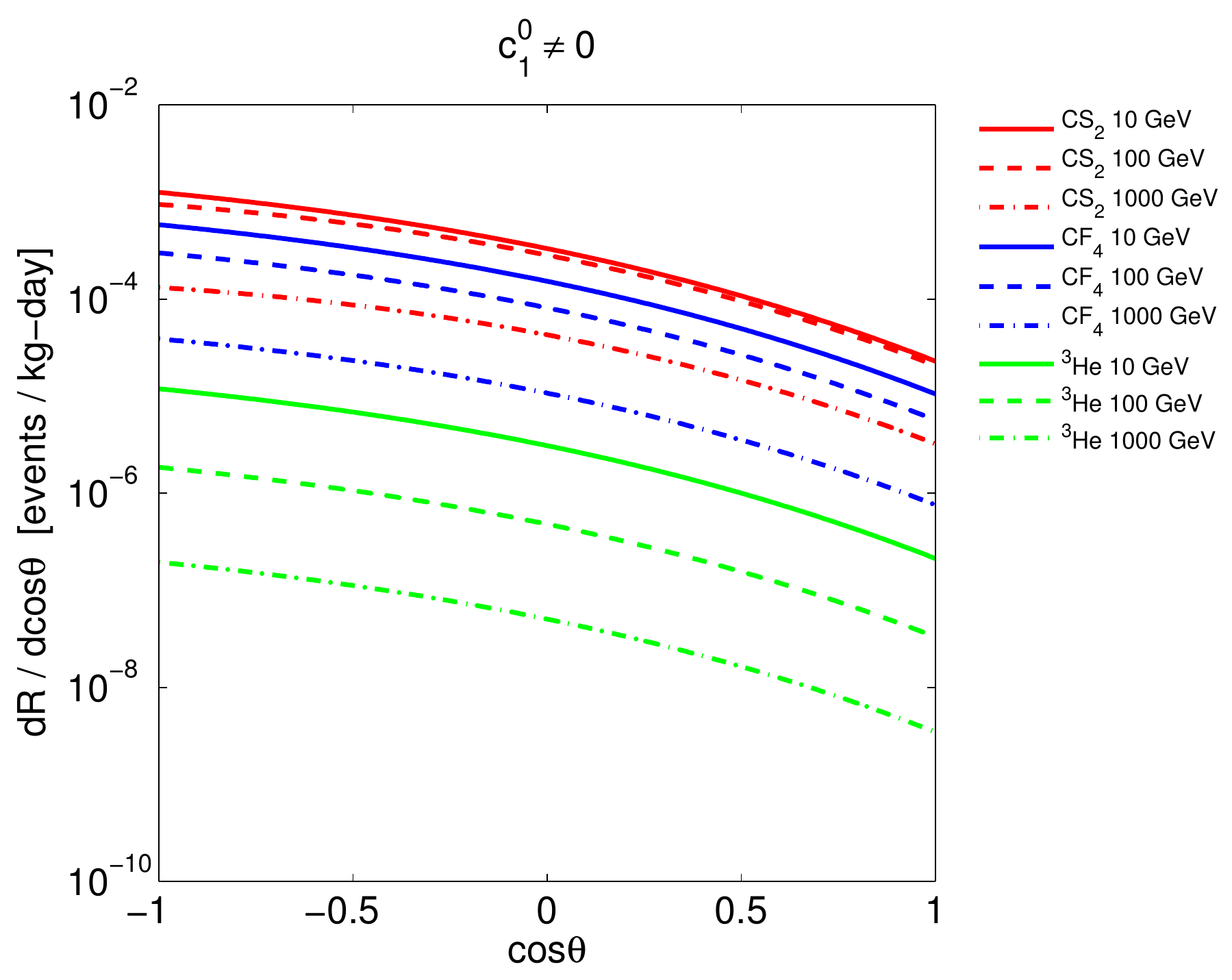}
\end{minipage}
\begin{minipage}[t]{0.497\linewidth}
\centering
\includegraphics[width=\textwidth]{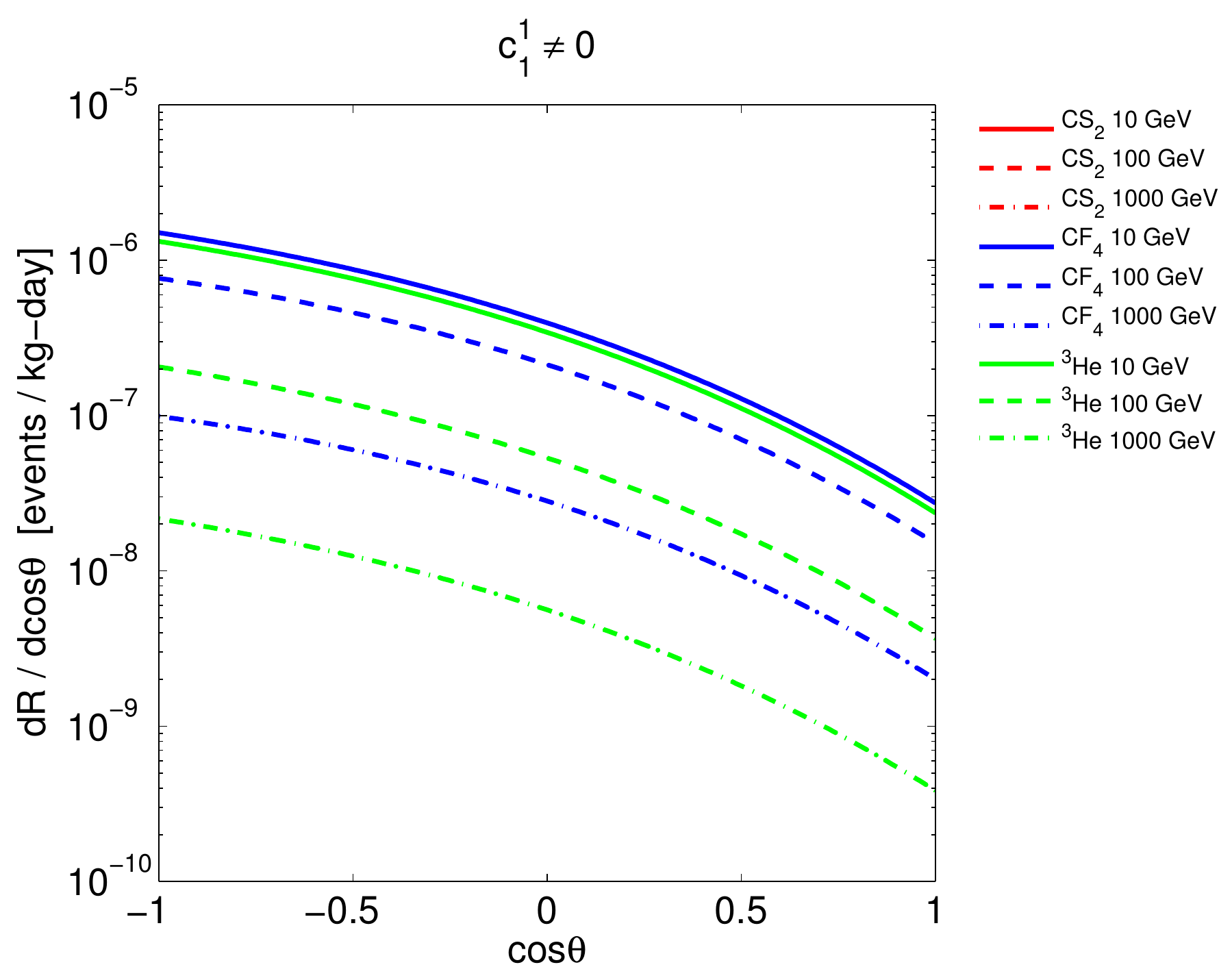}
\end{minipage}
\begin{minipage}[t]{0.49\linewidth}
\centering
\includegraphics[width=\textwidth]{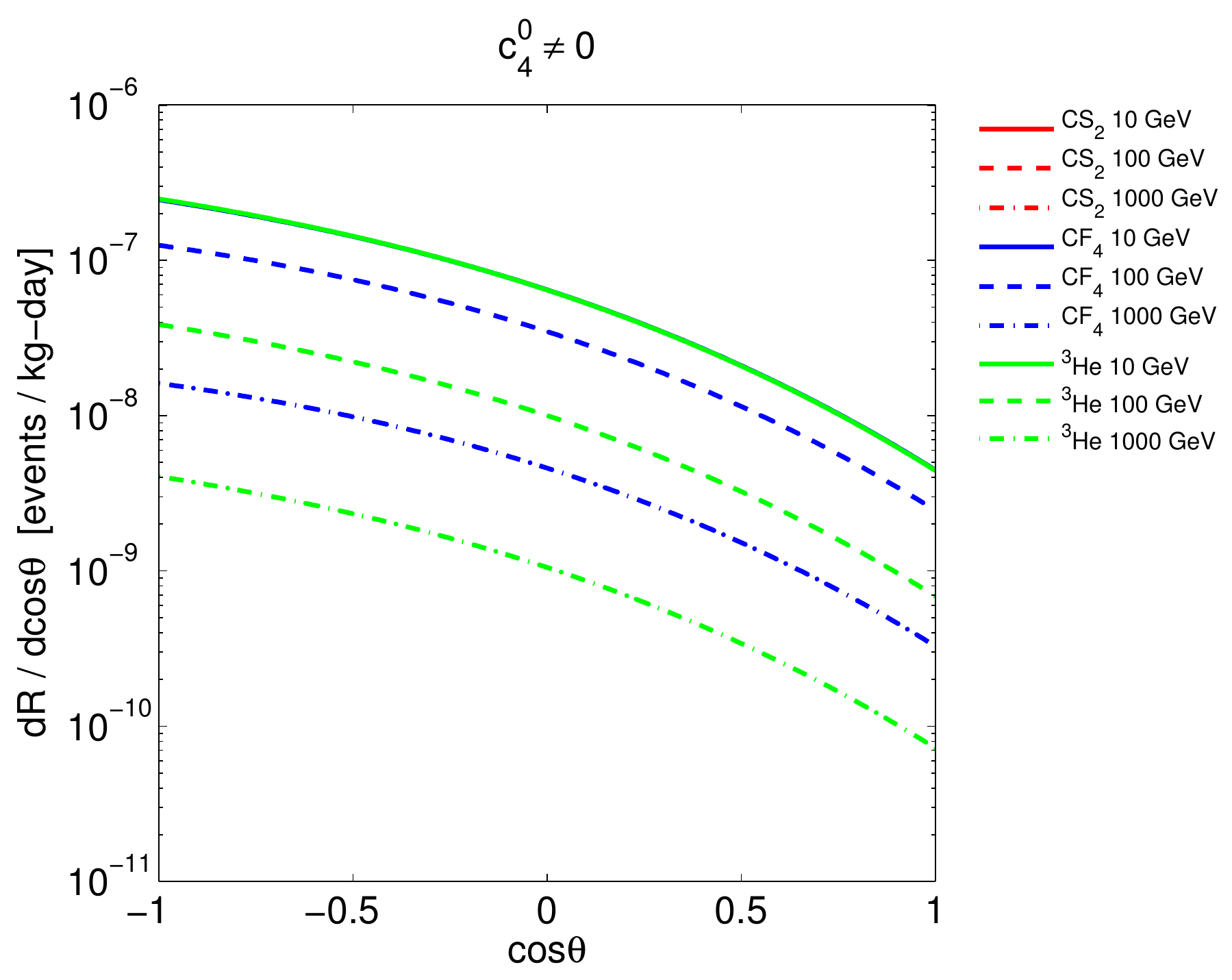}
\end{minipage}
\begin{minipage}[t]{0.49\linewidth}
\centering
\includegraphics[width=\textwidth]{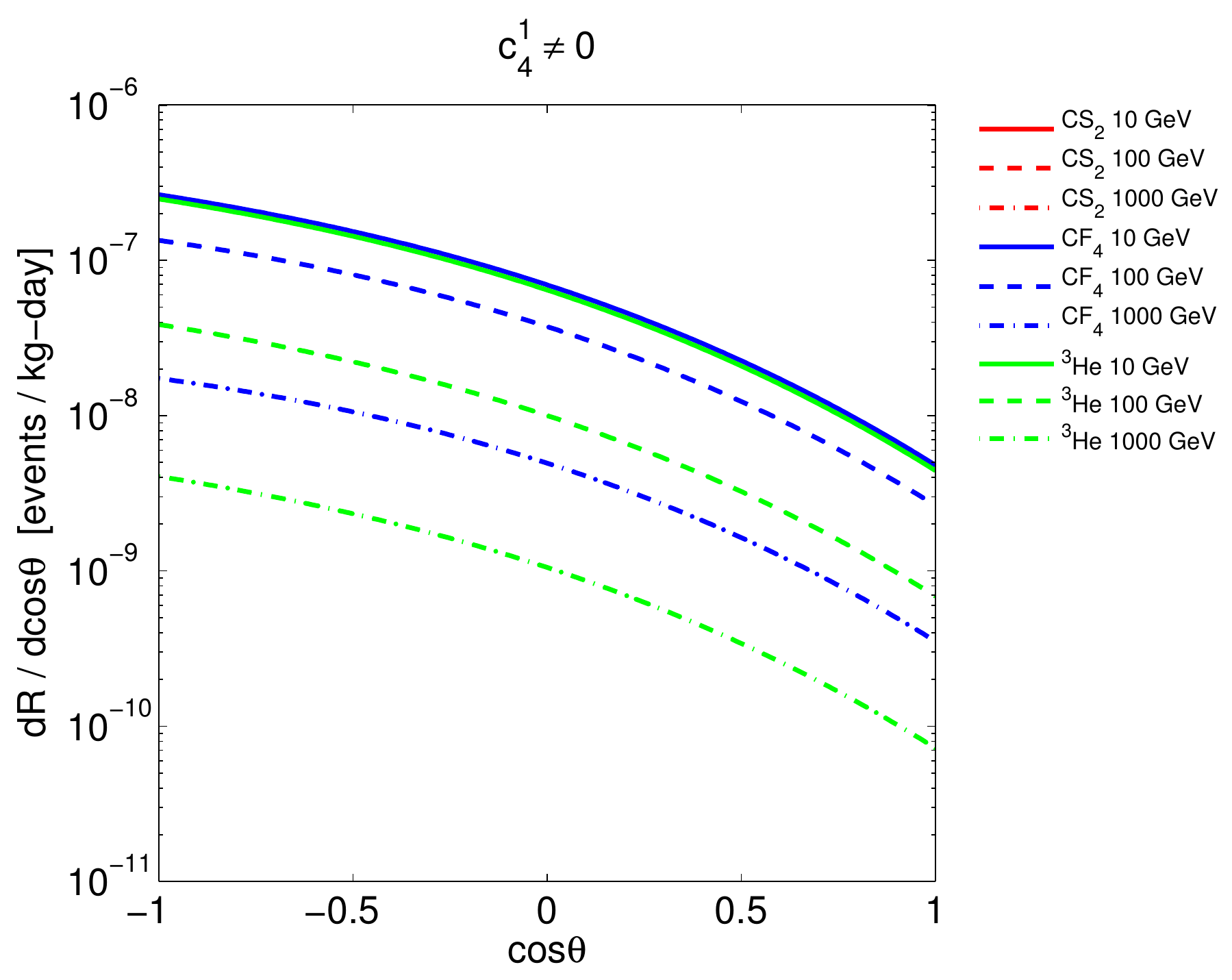}
\end{minipage}
\end{center}
\caption{Differential number of recoil events around $\cos\theta$ per unit time and per unit detector mass as a function of $\cos\theta$, i.e. ${\rm d}R/{\rm d}\hspace{-0.5mm}\cos\theta$ in Eq.~(\ref{eq:dR}). We report ${\rm d}R/{\rm d}\hspace{-0.5mm}\cos\theta$ for directional detectors with head-tail discrimination made of CS$_2$, CF$_4$ and $^3$He. Solid, dashed and dot-dashed lines correspond to $m_\chi$ equal to 10 GeV, 100 GeV and 1000 GeV, respectively, as shown in the legends. The top panels refer to the coupling constants $c_1^0$ and $c_1^1$, whereas the bottom panels correspond to $c_4^0$ and $c_4^1$, respectively. Here we assume a truncated Maxwell-Boltzmann distribution for the Milky Way dark matter particles.}
\label{fig:c1c4}
\end{figure}
\begin{figure}[t]
\begin{center}
\begin{minipage}[t]{0.49\linewidth}
\centering
\includegraphics[width=\textwidth]{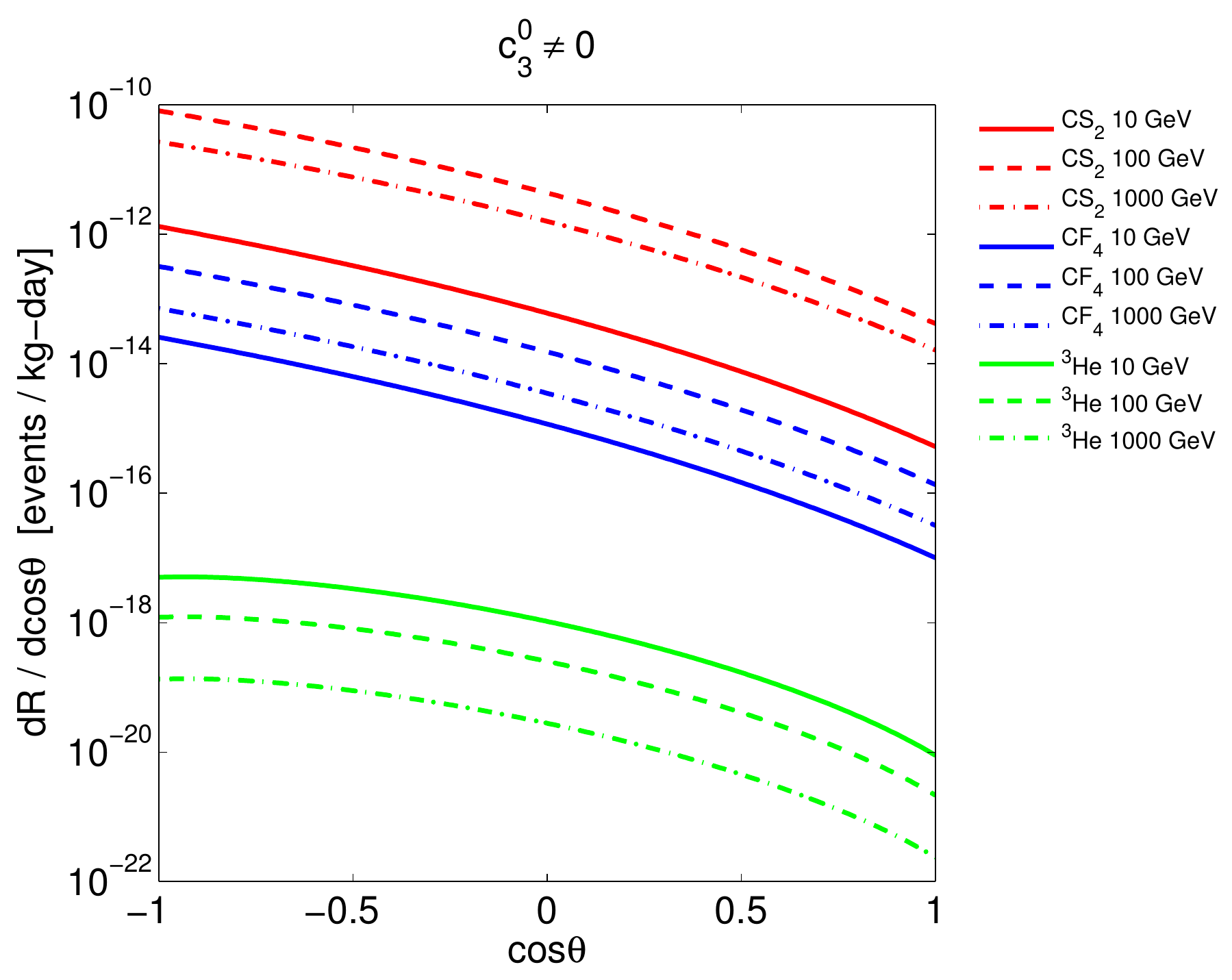}
\end{minipage}
\begin{minipage}[t]{0.497\linewidth}
\centering
\includegraphics[width=\textwidth]{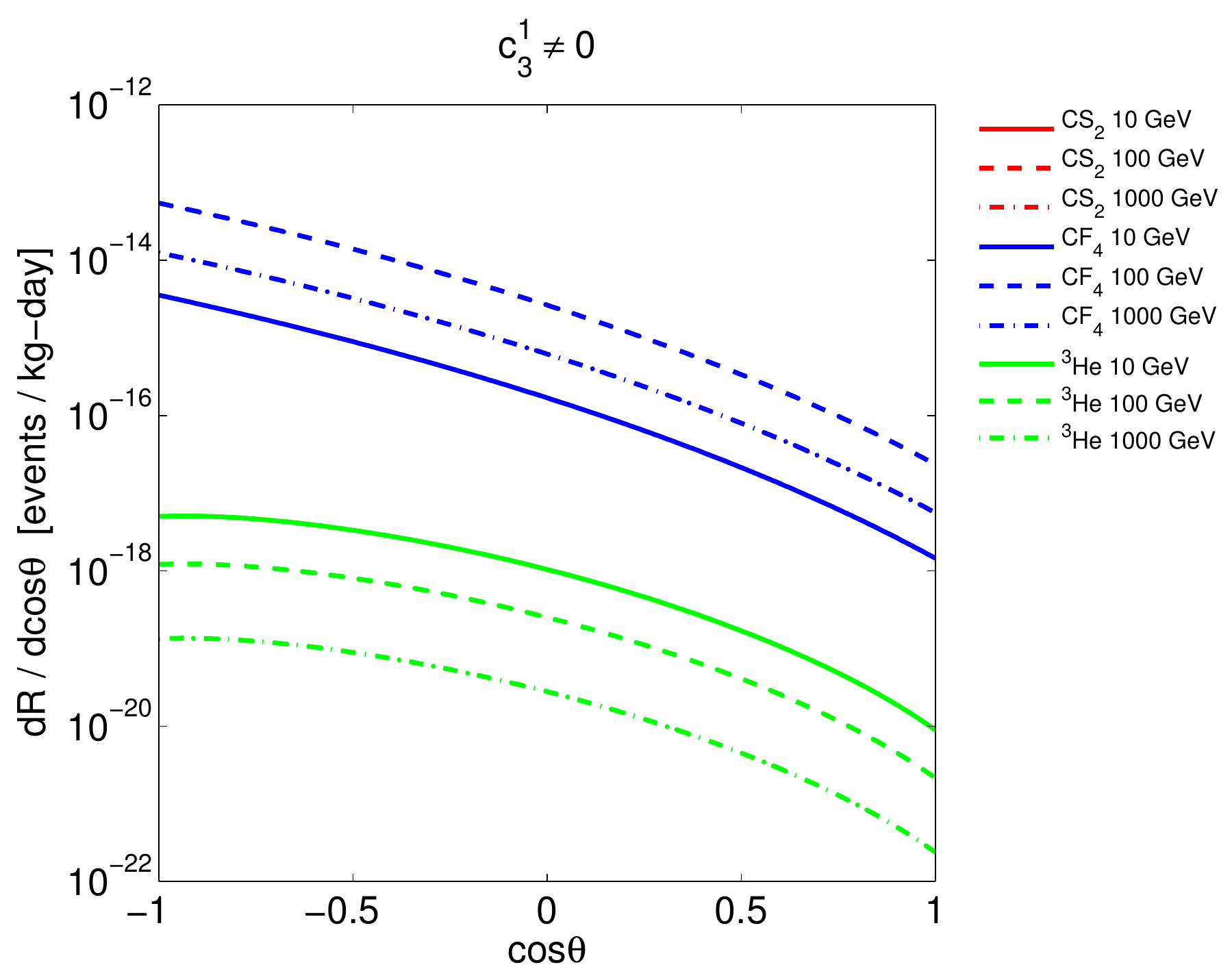}
\end{minipage}
\begin{minipage}[t]{0.49\linewidth}
\centering
\includegraphics[width=\textwidth]{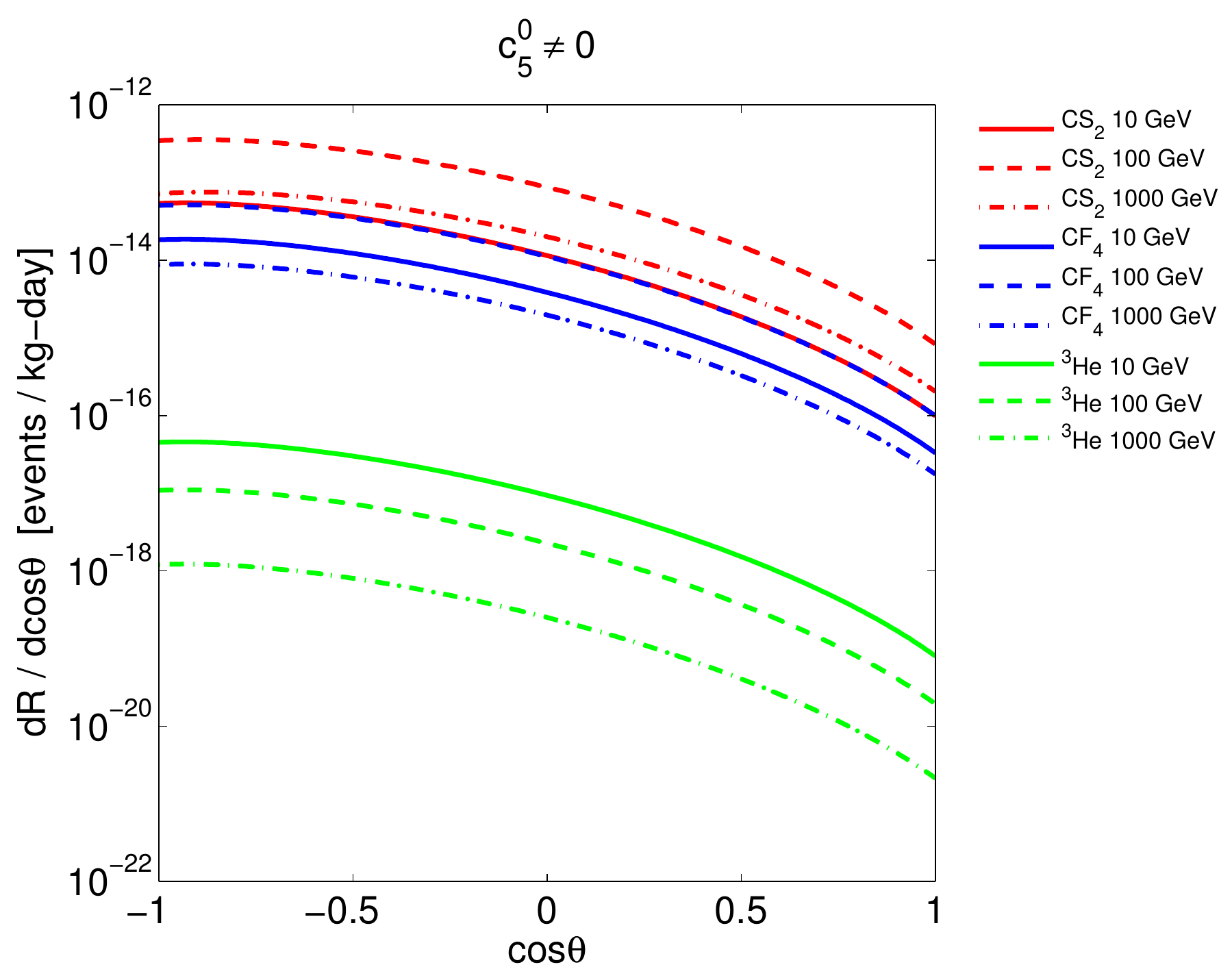}
\end{minipage}
\begin{minipage}[t]{0.49\linewidth}
\centering
\includegraphics[width=\textwidth]{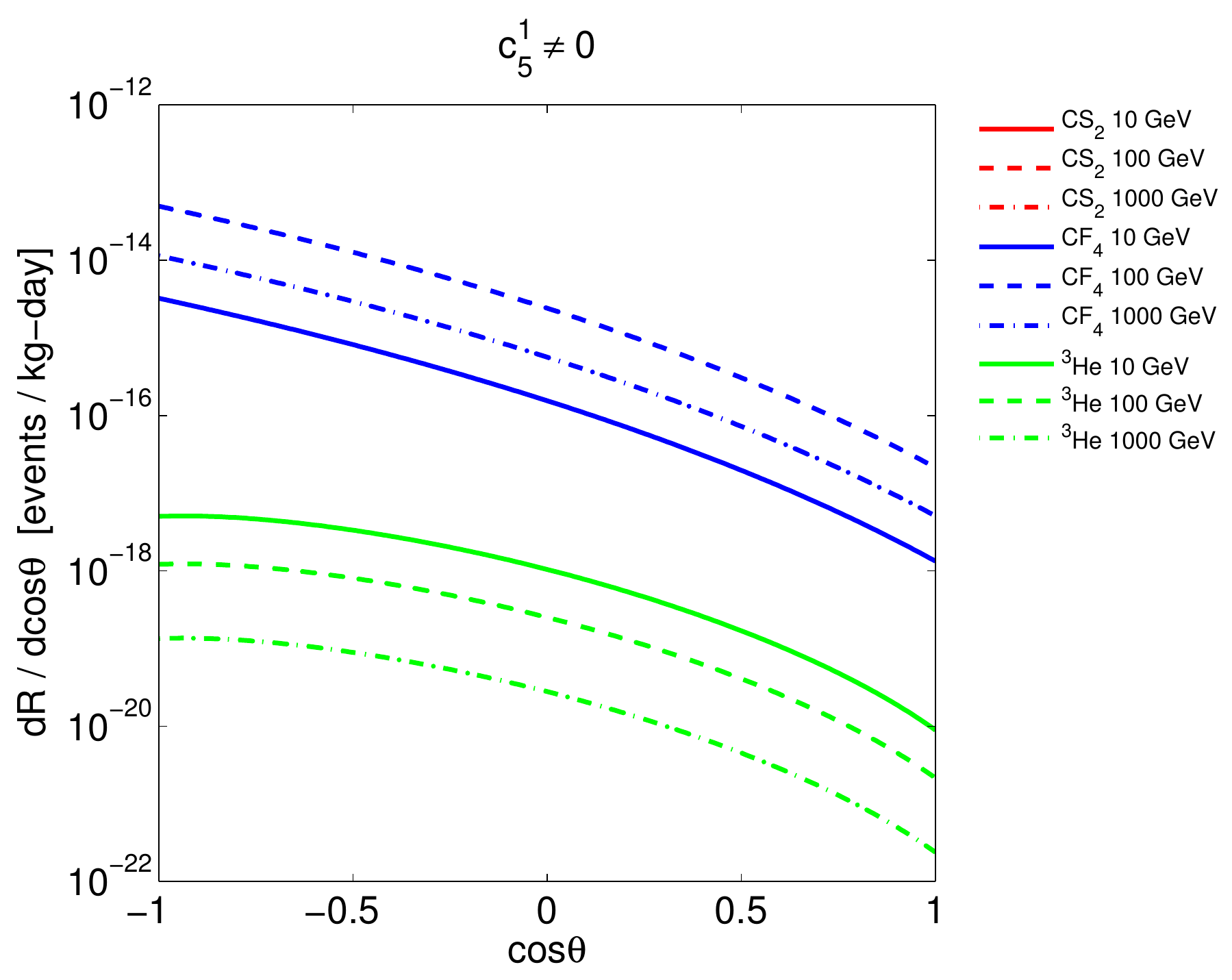}
\end{minipage}
\end{center}
\caption{Same as Fig.~\ref{fig:c1c4}, but now for the operators $\hat{\mathcal{O}}_3$ and $\hat{\mathcal{O}}_5$.}
\label{fig:c3c5}
\end{figure} 
\begin{figure}[t]
\begin{center}
\begin{minipage}[t]{0.49\linewidth}
\centering
\includegraphics[width=\textwidth]{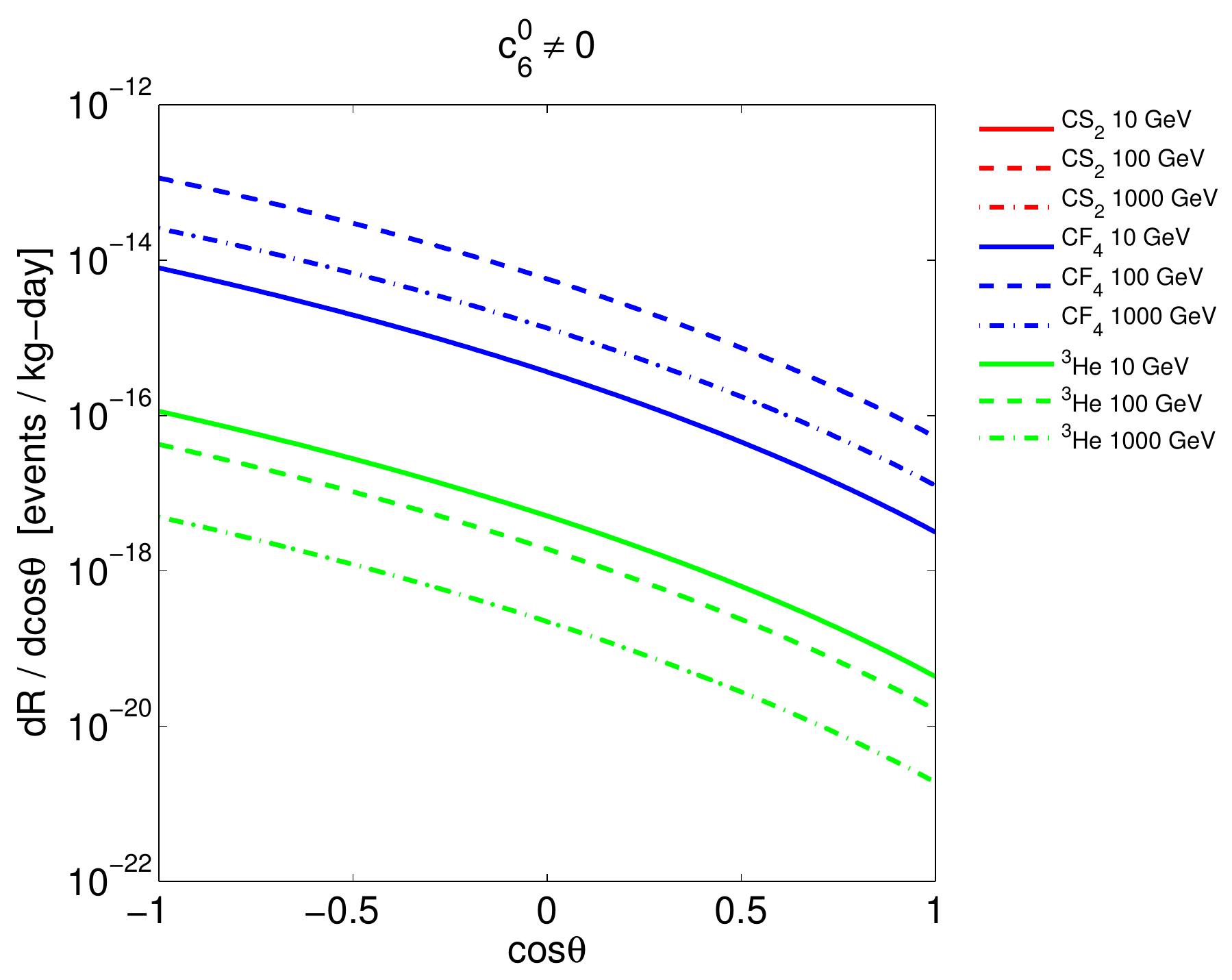}
\end{minipage}
\begin{minipage}[t]{0.497\linewidth}
\centering
\includegraphics[width=\textwidth]{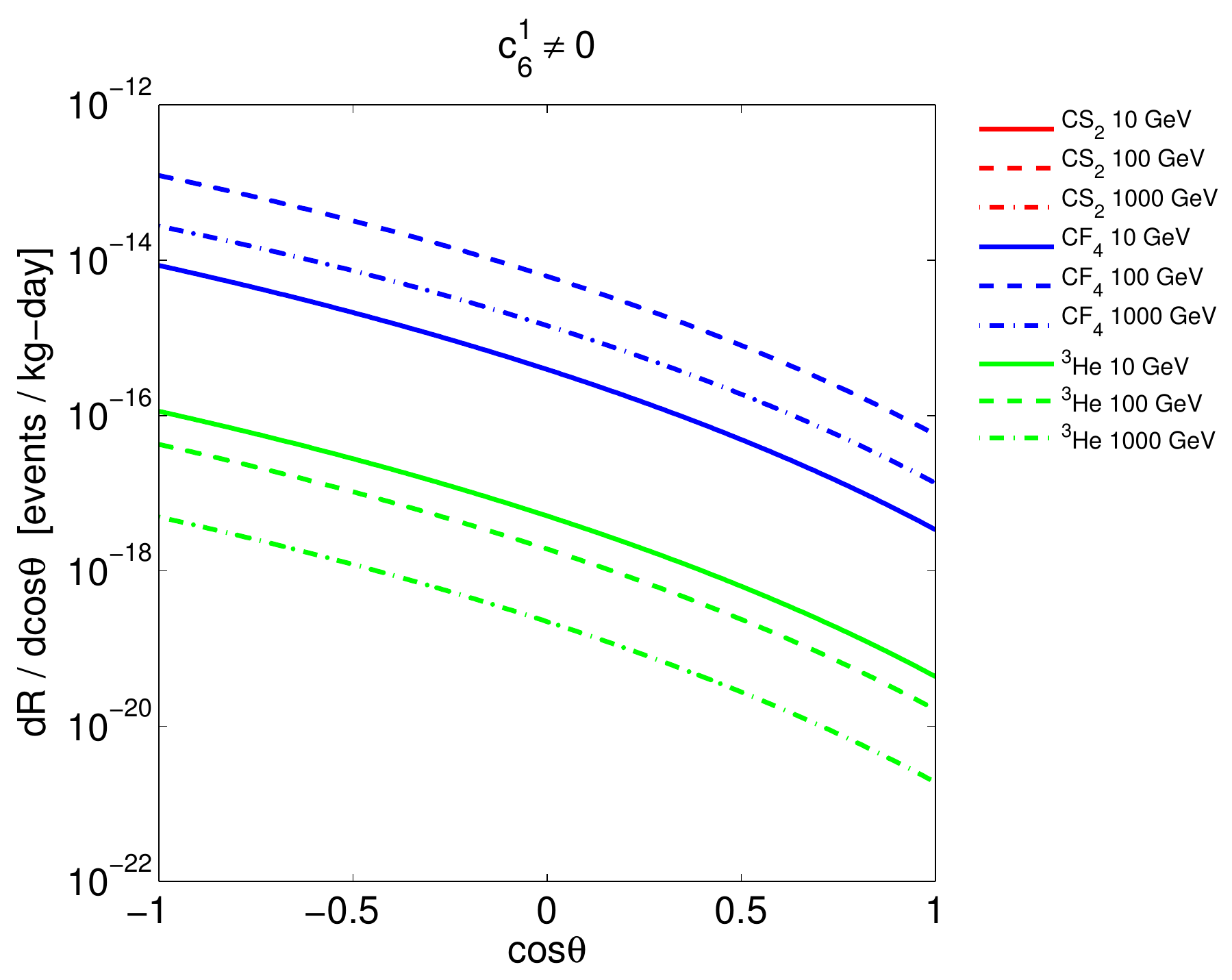}
\end{minipage}
\begin{minipage}[t]{0.49\linewidth}
\centering
\includegraphics[width=\textwidth]{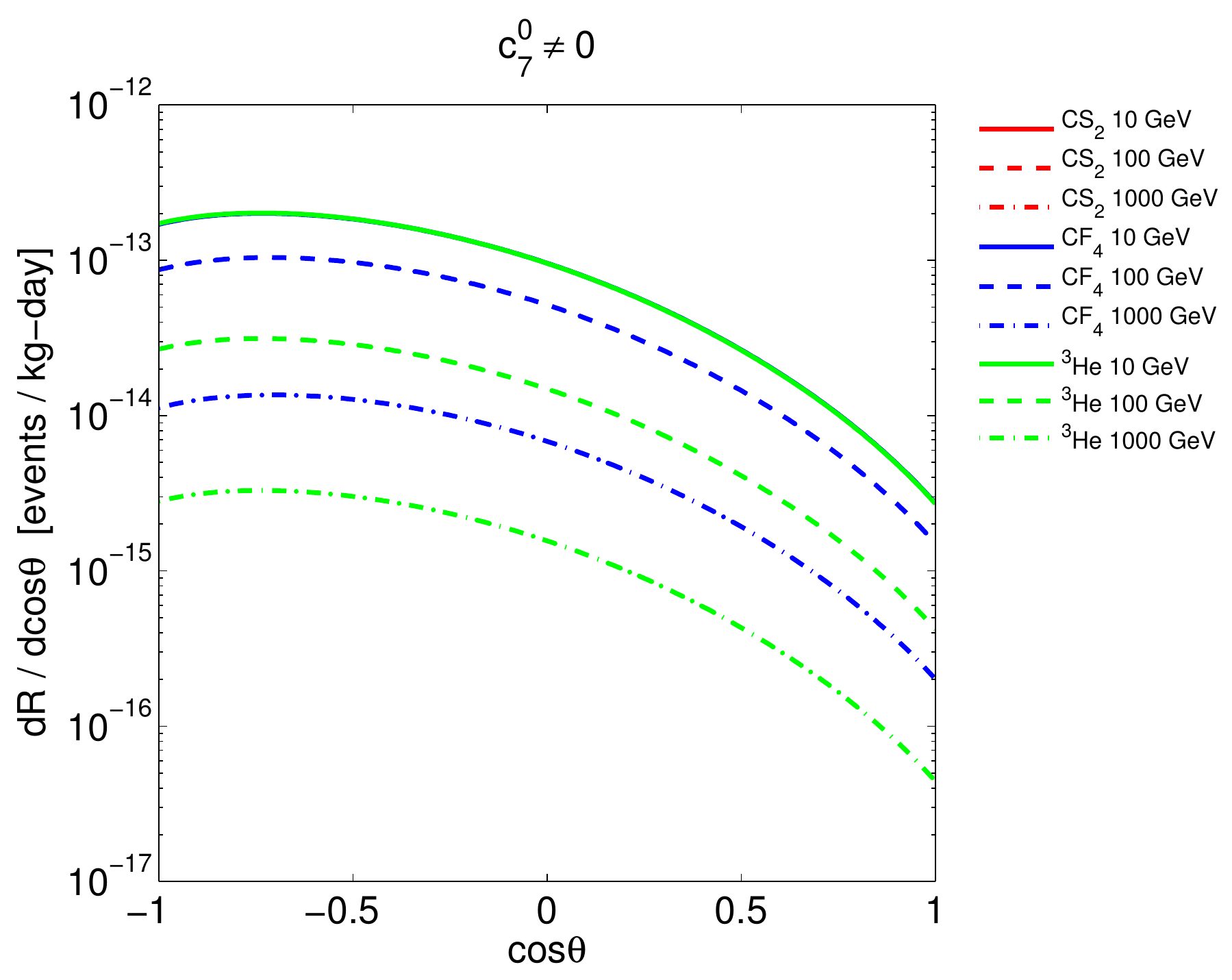}
\end{minipage}
\begin{minipage}[t]{0.49\linewidth}
\centering
\includegraphics[width=\textwidth]{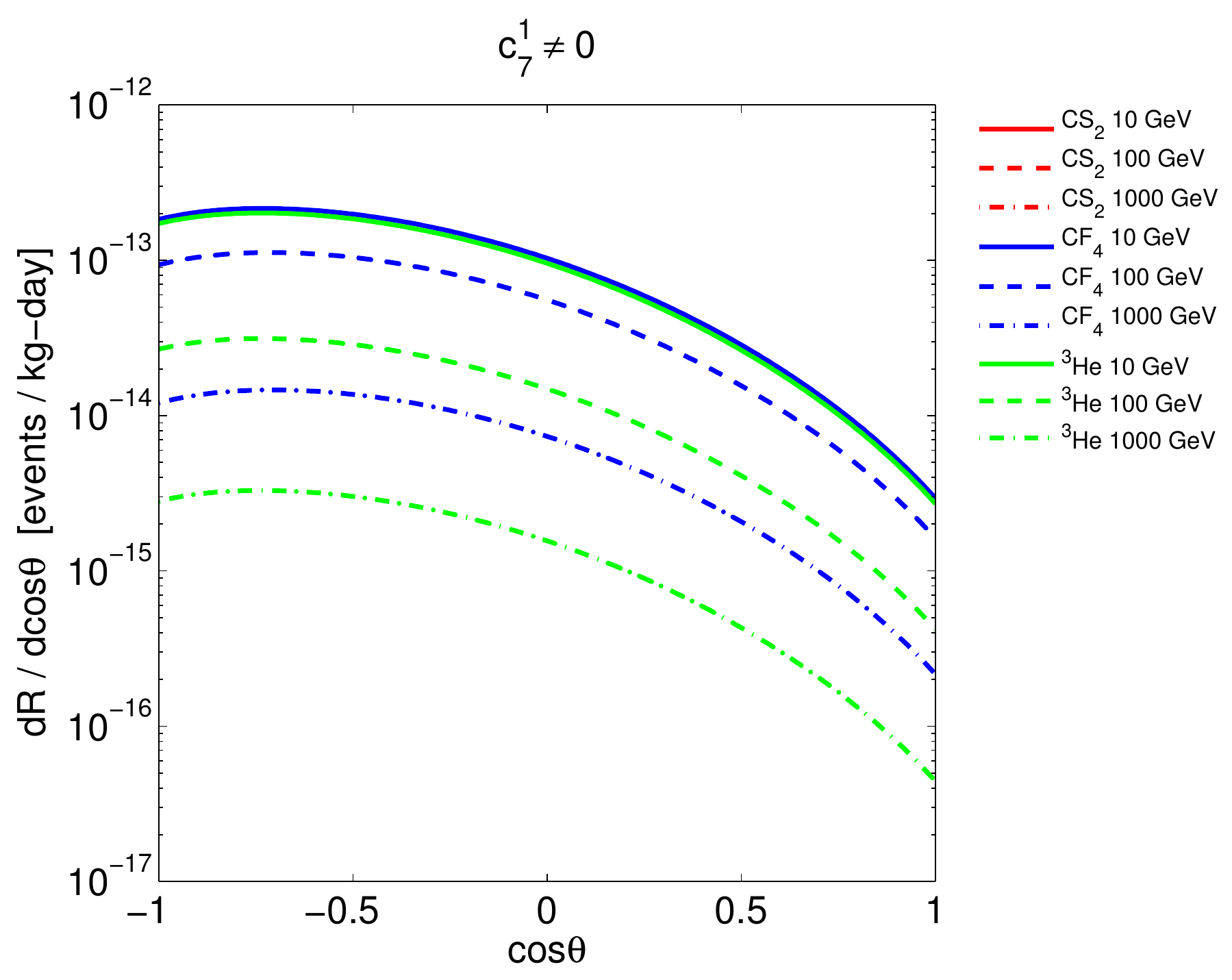}
\end{minipage}
\end{center}
\caption{Same as Fig.~\ref{fig:c1c4}, but now for the operators $\hat{\mathcal{O}}_6$ and $\hat{\mathcal{O}}_7$.}
\label{fig:c6c7}
\end{figure}
\begin{figure}[t]
\begin{center}
\begin{minipage}[t]{0.49\linewidth}
\centering
\includegraphics[width=\textwidth]{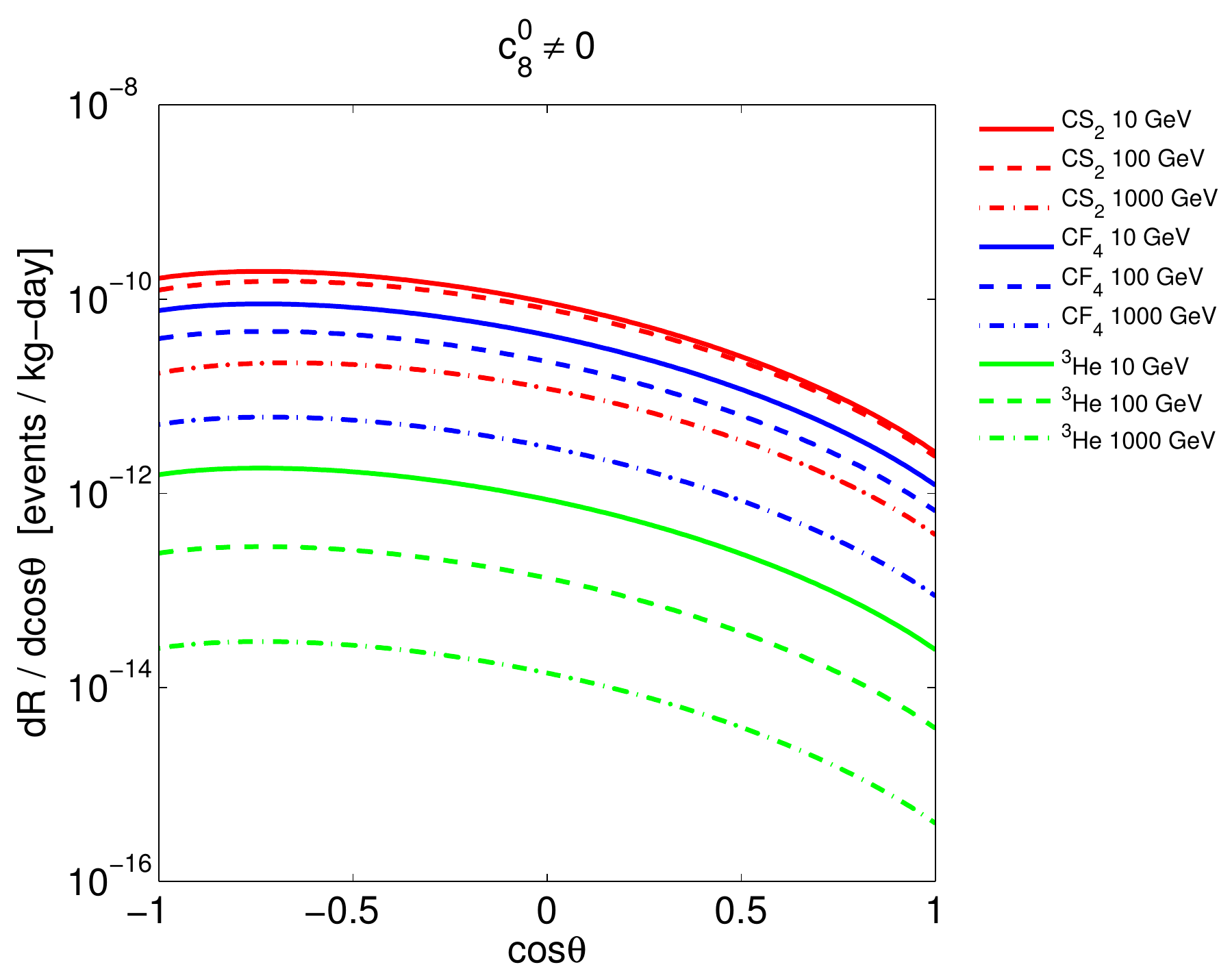}
\end{minipage}
\begin{minipage}[t]{0.497\linewidth}
\centering
\includegraphics[width=\textwidth]{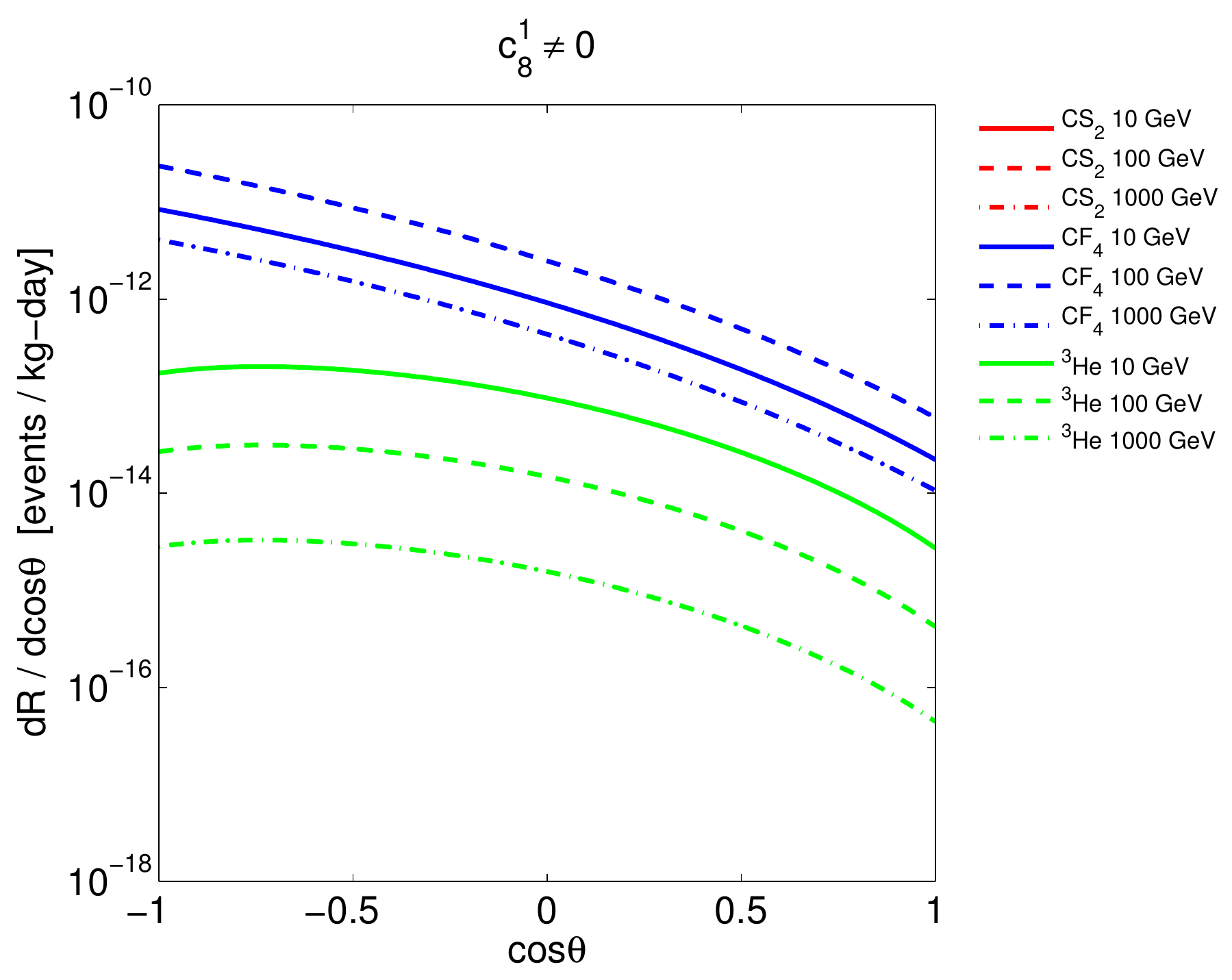}
\end{minipage}
\begin{minipage}[t]{0.49\linewidth}
\centering
\includegraphics[width=\textwidth]{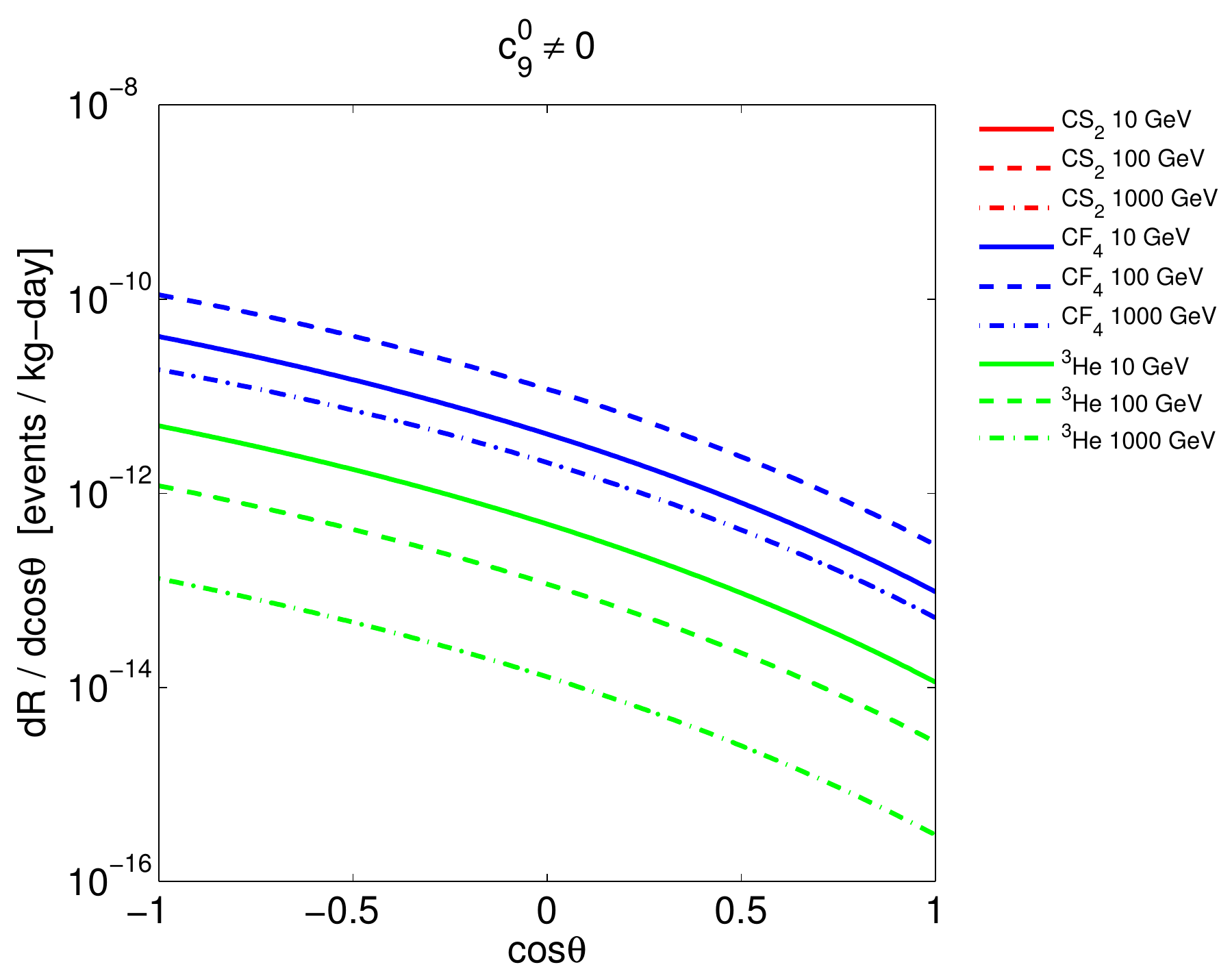}
\end{minipage}
\begin{minipage}[t]{0.49\linewidth}
\centering
\includegraphics[width=\textwidth]{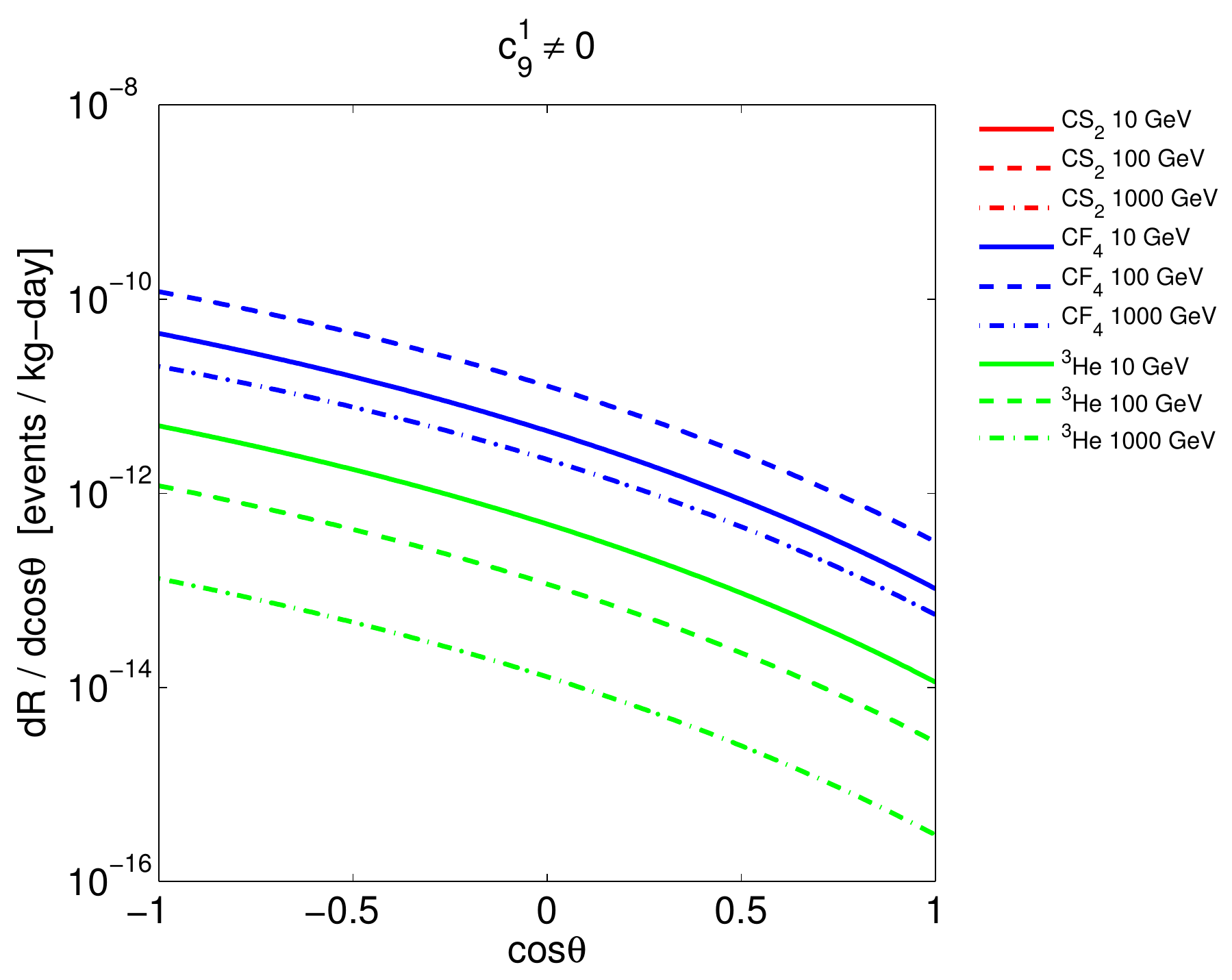}
\end{minipage}
\end{center}
\caption{Same as Fig.~\ref{fig:c1c4}, but now for the operators $\hat{\mathcal{O}}_8$ and $\hat{\mathcal{O}}_9$.}
\label{fig:c8c9}
\end{figure}
\begin{figure}[t]
\begin{center}
\begin{minipage}[t]{0.49\linewidth}
\centering
\includegraphics[width=\textwidth]{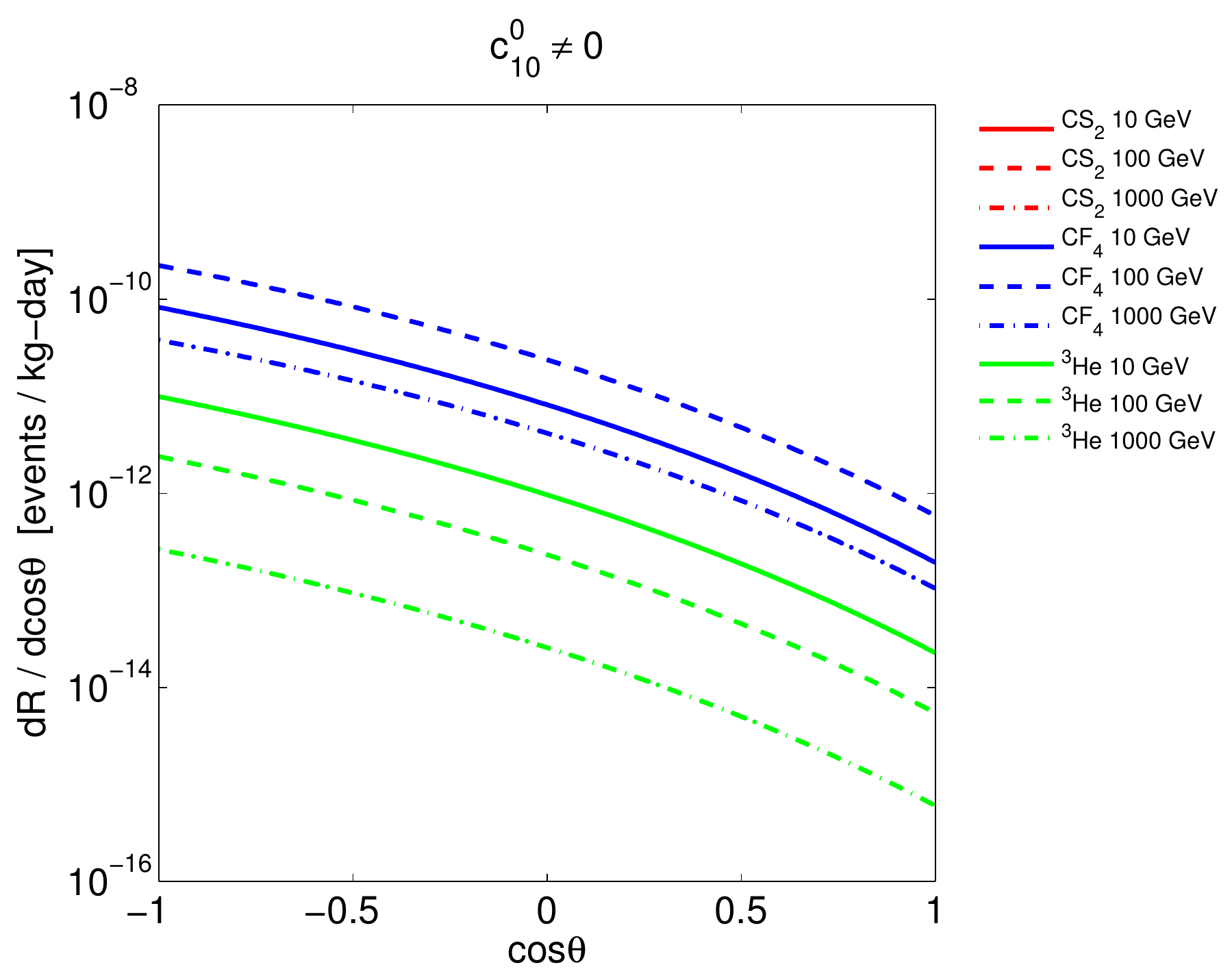}
\end{minipage}
\begin{minipage}[t]{0.497\linewidth}
\centering
\includegraphics[width=\textwidth]{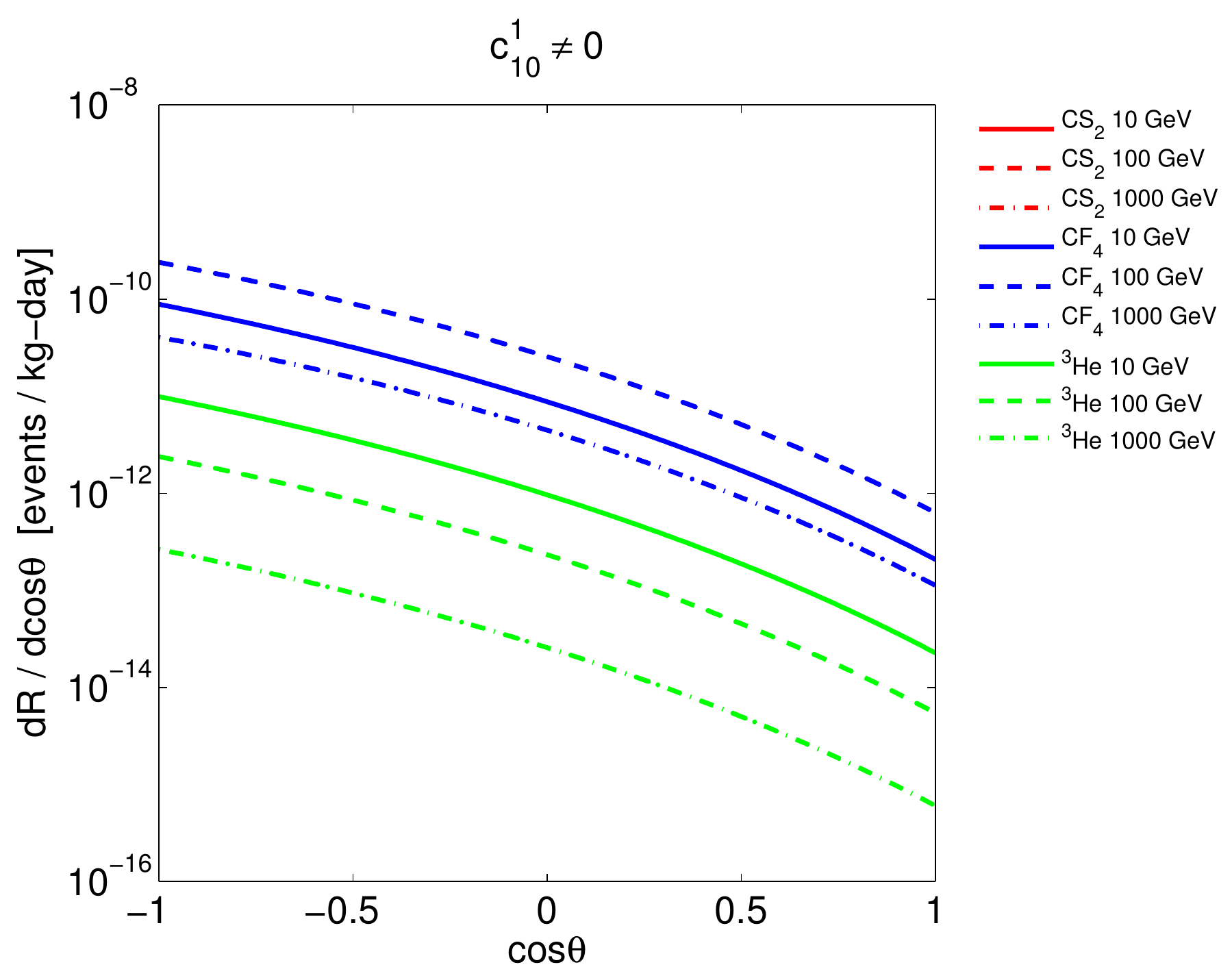}
\end{minipage}
\begin{minipage}[t]{0.49\linewidth}
\centering
\includegraphics[width=\textwidth]{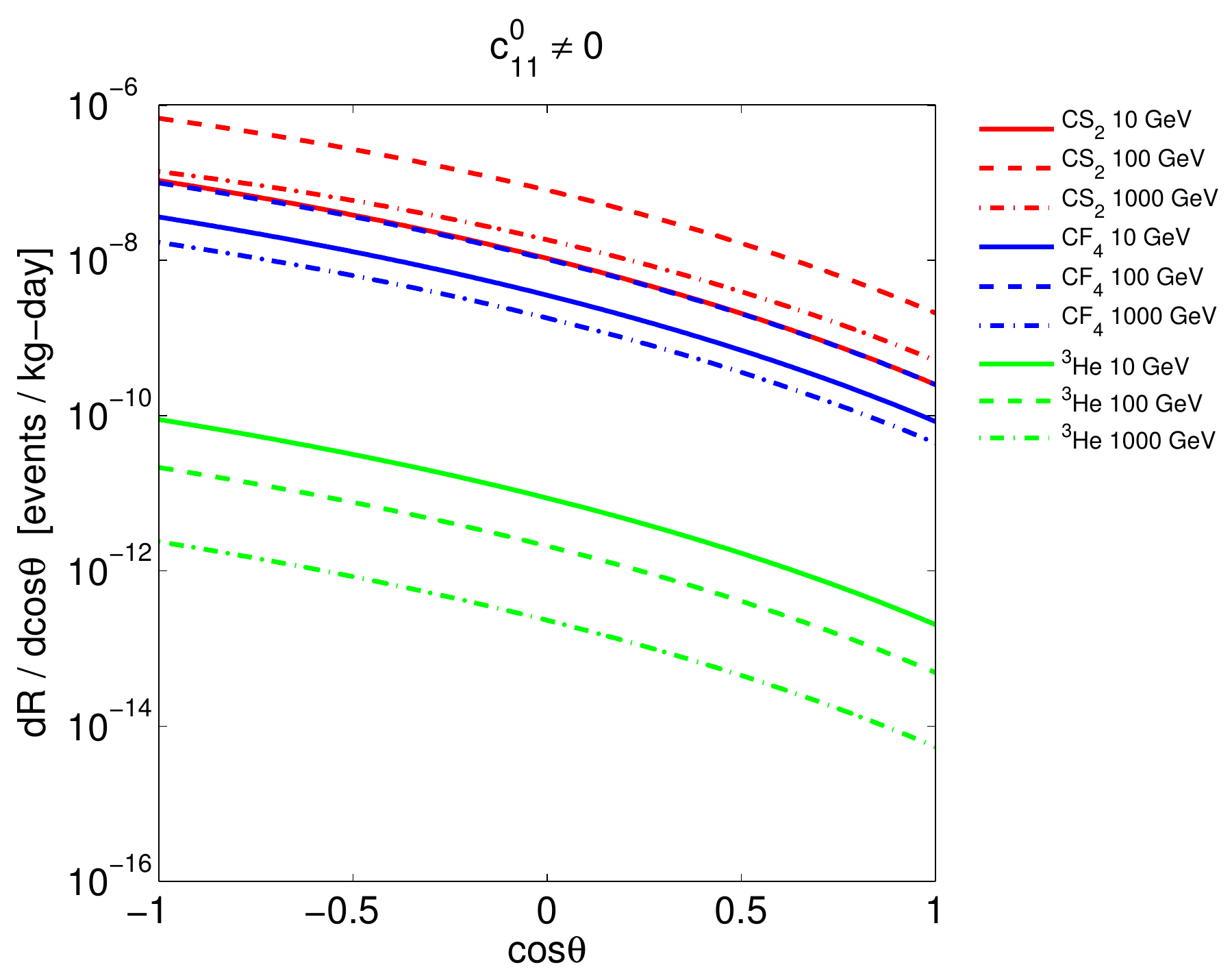}
\end{minipage}
\begin{minipage}[t]{0.49\linewidth}
\centering
\includegraphics[width=\textwidth]{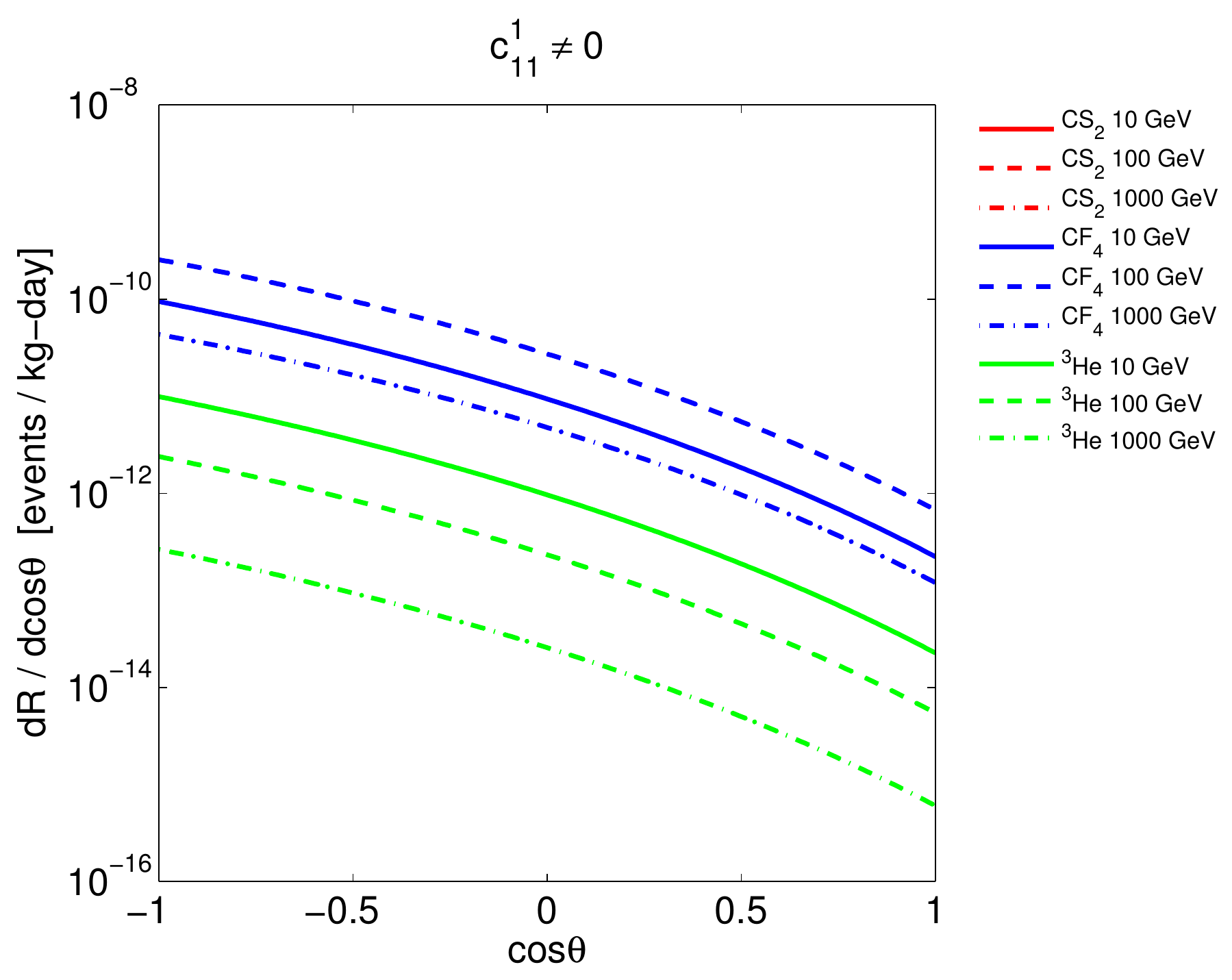}
\end{minipage}
\end{center}
\caption{Same as Fig.~\ref{fig:c1c4}, but now for the operators $\hat{\mathcal{O}}_{10}$ and $\hat{\mathcal{O}}_{11}$.}
\label{fig:c10c11}
\end{figure}
\begin{figure}[t]
\begin{center}
\begin{minipage}[t]{0.49\linewidth}
\centering
\includegraphics[width=\textwidth]{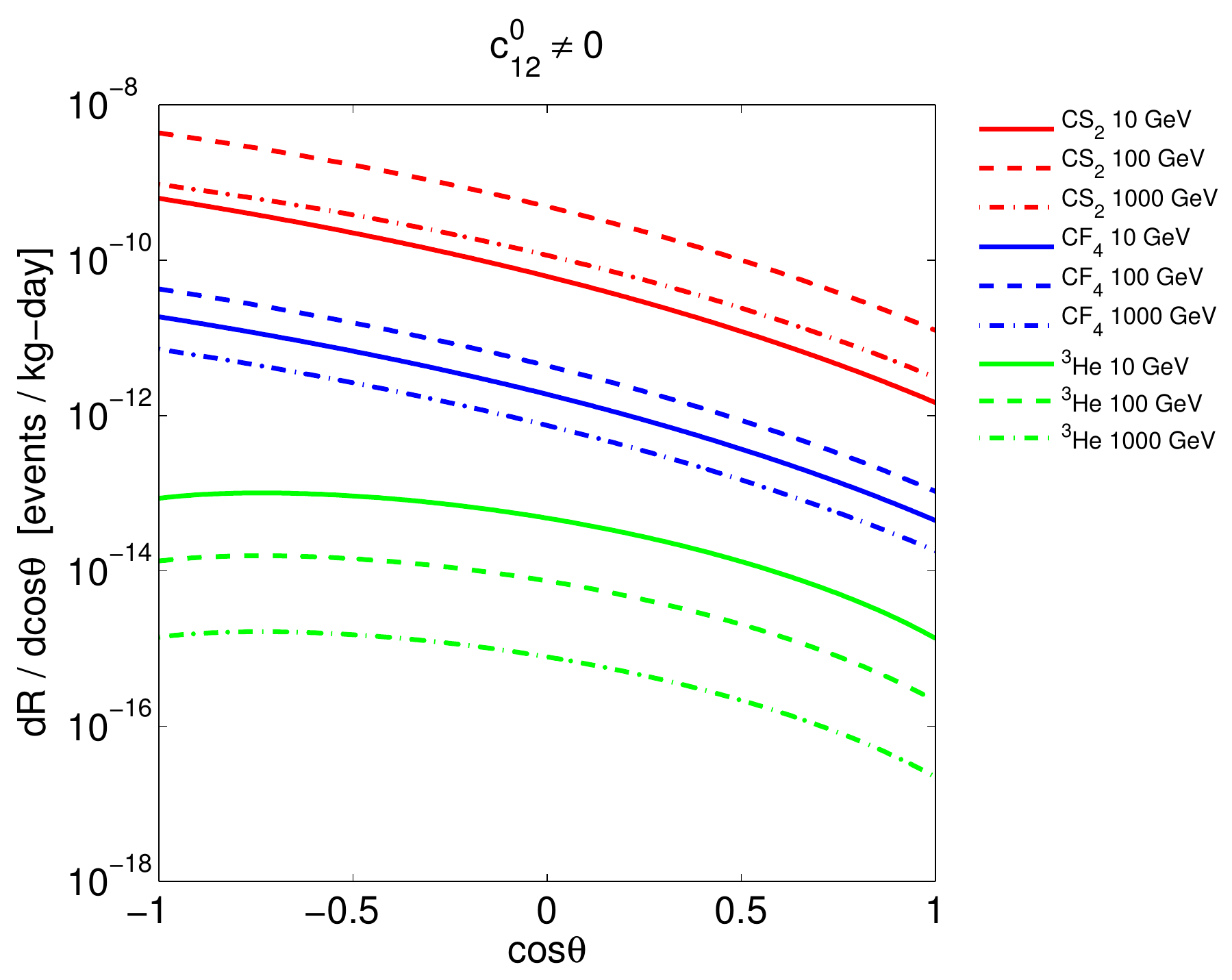}
\end{minipage}
\begin{minipage}[t]{0.497\linewidth}
\centering
\includegraphics[width=\textwidth]{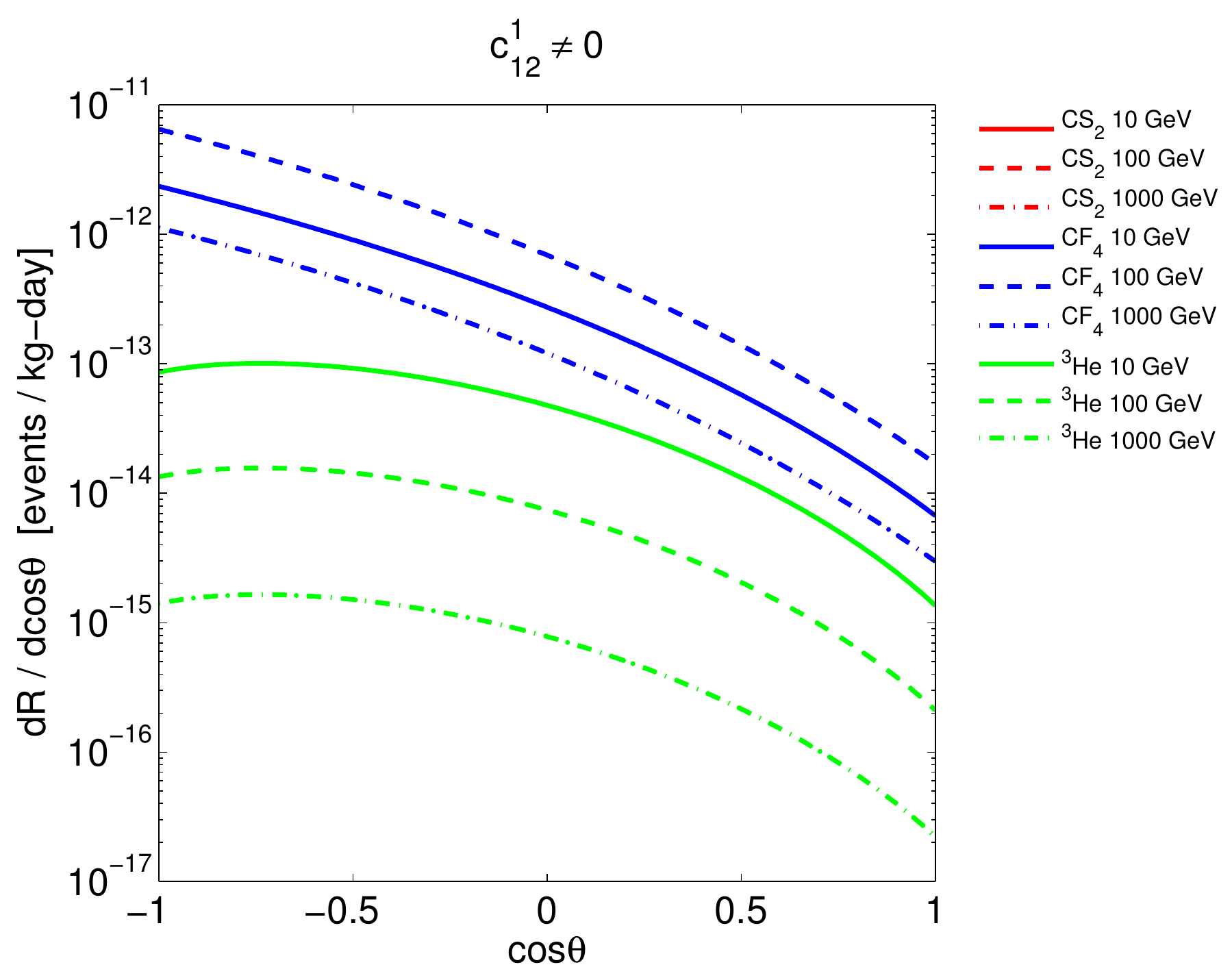}
\end{minipage}
\begin{minipage}[t]{0.49\linewidth}
\centering
\includegraphics[width=\textwidth]{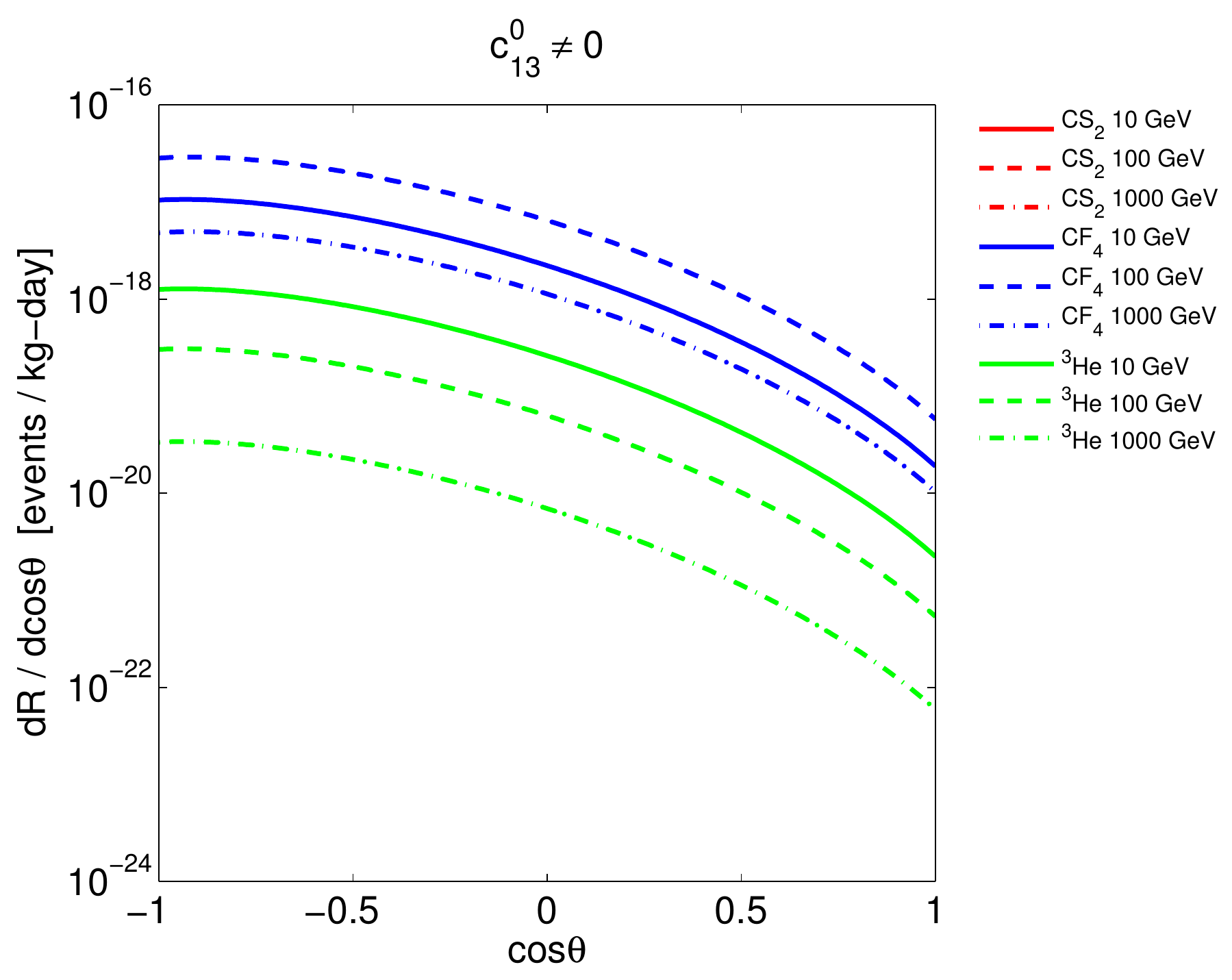}
\end{minipage}
\begin{minipage}[t]{0.49\linewidth}
\centering
\includegraphics[width=\textwidth]{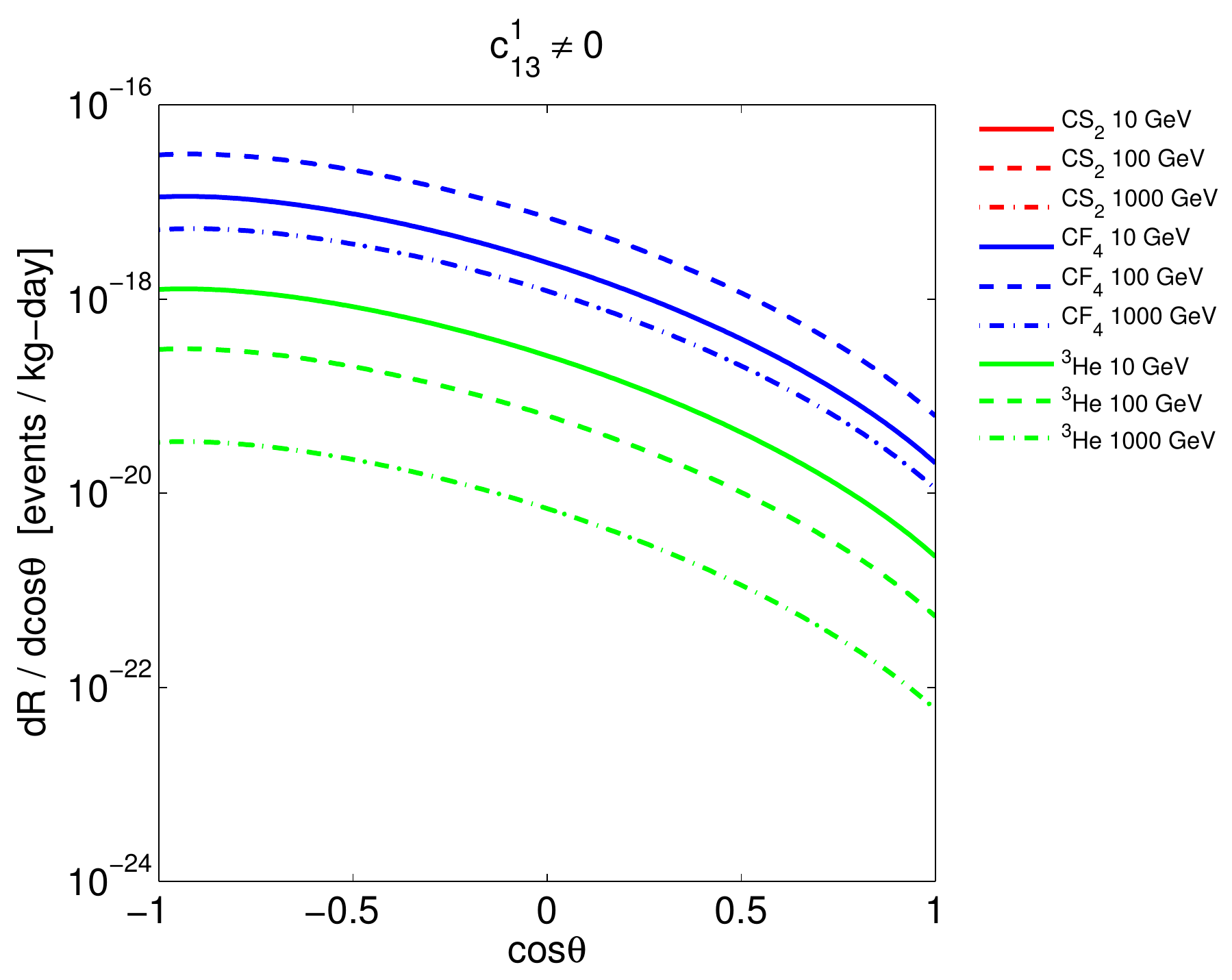}
\end{minipage}
\end{center}
\caption{Same as Fig.~\ref{fig:c1c4}, but now for the operators $\hat{\mathcal{O}}_{12}$ and $\hat{\mathcal{O}}_{13}$.}
\label{fig:c12c13}
\end{figure}
\begin{figure}[t]
\begin{center}
\begin{minipage}[t]{0.49\linewidth}
\centering
\includegraphics[width=\textwidth]{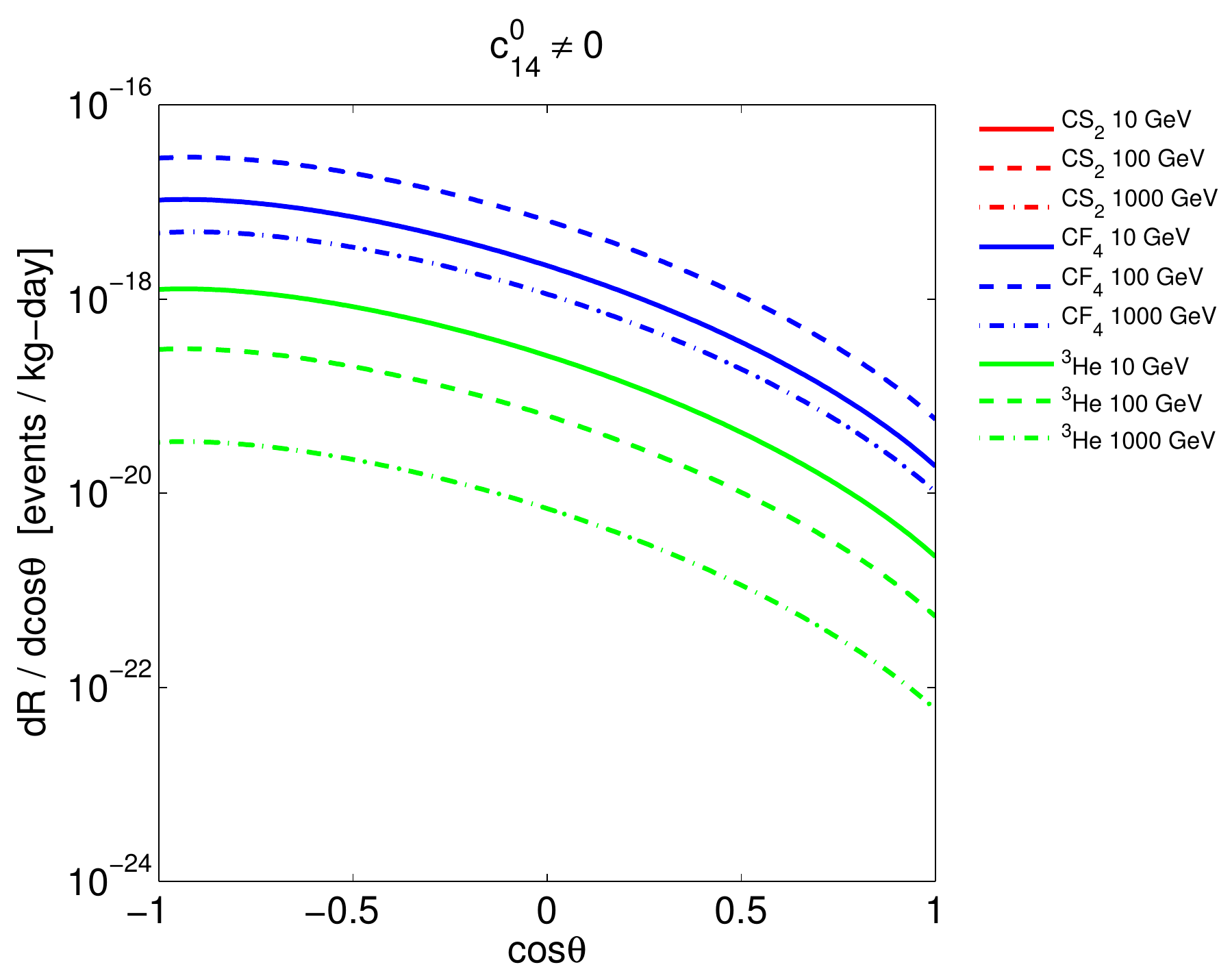}
\end{minipage}
\begin{minipage}[t]{0.497\linewidth}
\centering
\includegraphics[width=\textwidth]{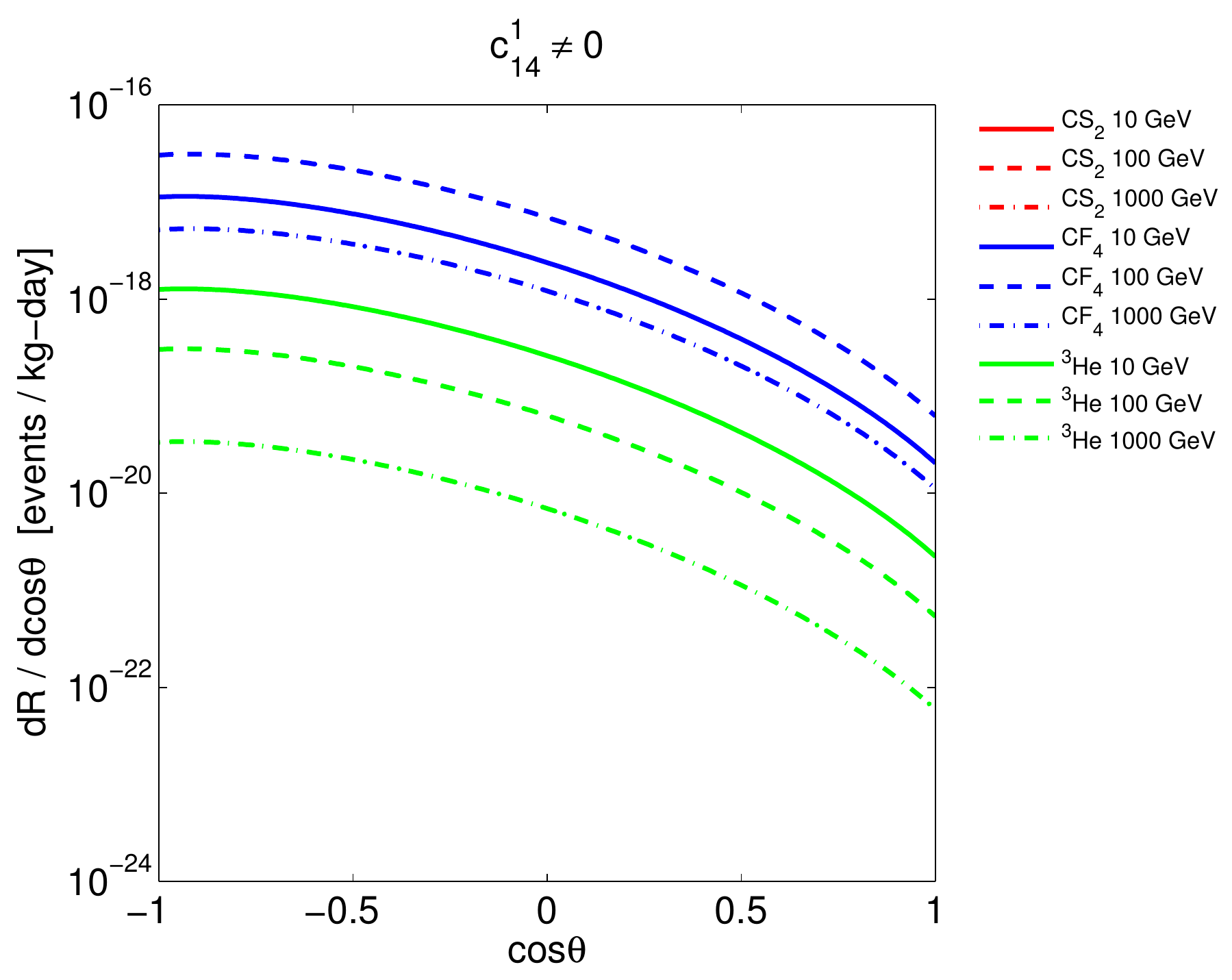}
\end{minipage}
\begin{minipage}[t]{0.49\linewidth}
\centering
\includegraphics[width=\textwidth]{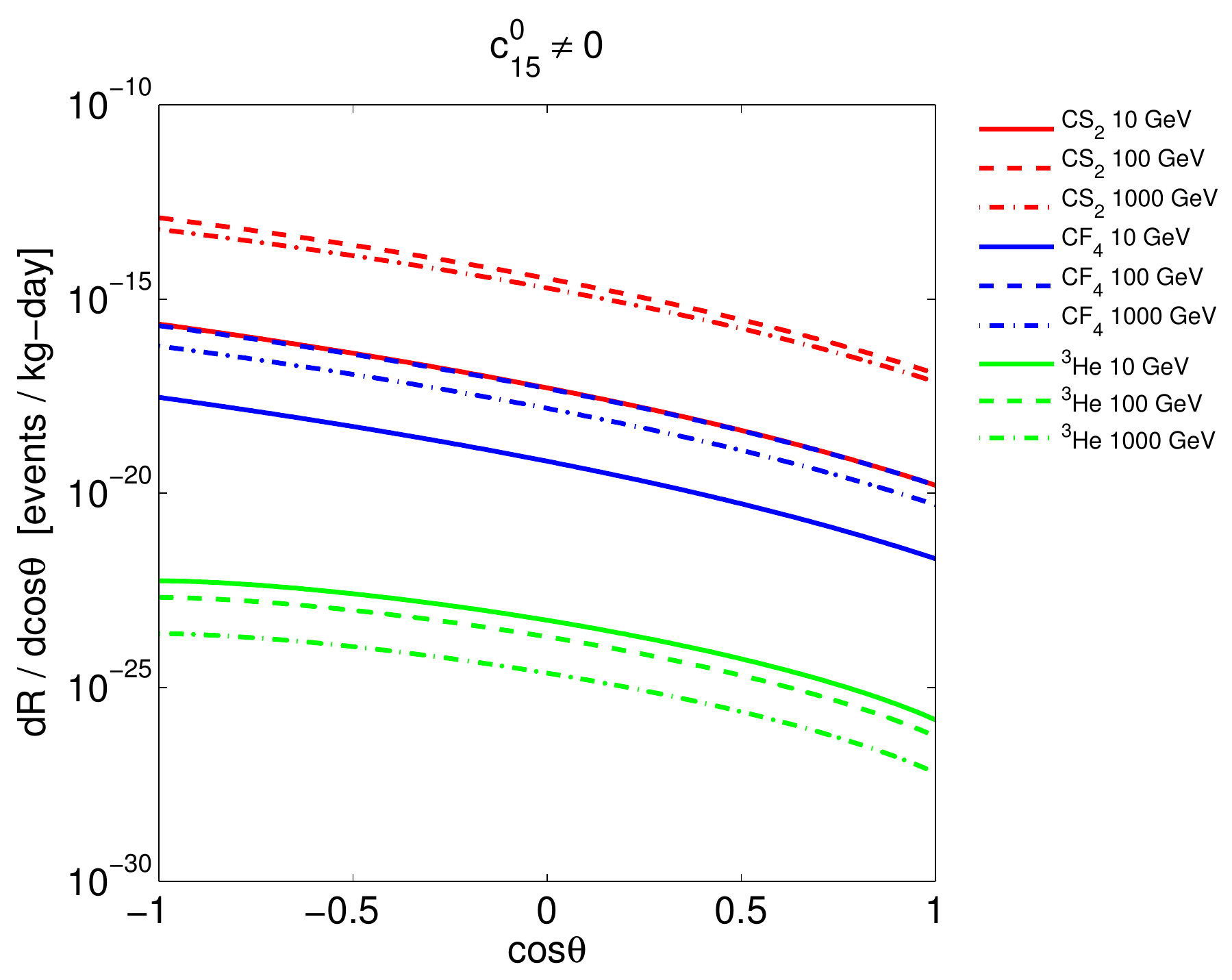}
\end{minipage}
\begin{minipage}[t]{0.49\linewidth}
\centering
\includegraphics[width=\textwidth]{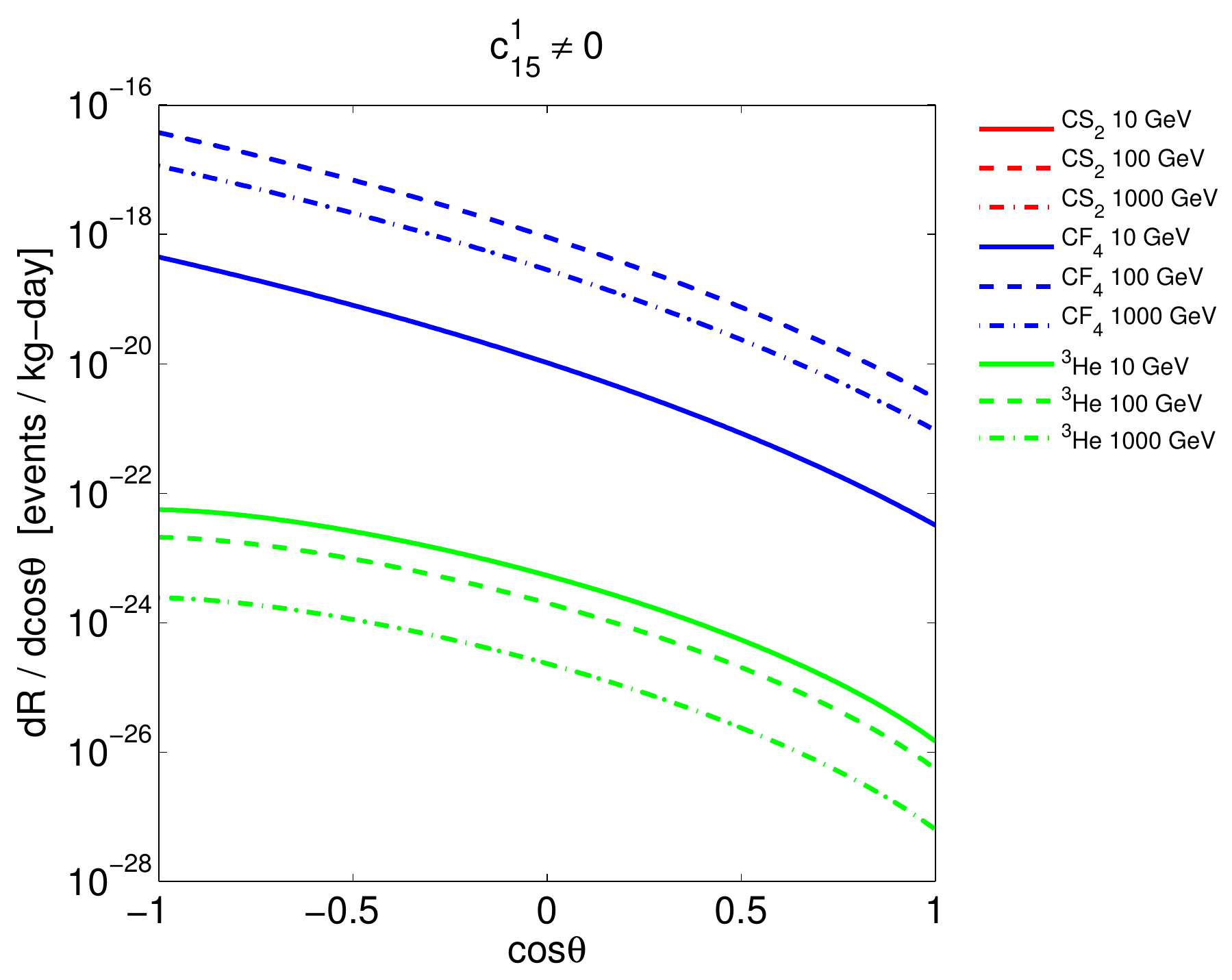}
\end{minipage}
\end{center}
\caption{Same as Fig.~\ref{fig:c1c4}, but now for the operators $\hat{\mathcal{O}}_{14}$ and $\hat{\mathcal{O}}_{15}$.}
\label{fig:c14c15}
\end{figure}

The Earth's velocity ${\bf v}_e(t)$ in the equations of Secs.~\ref{sec:Rate} and \ref{sec:Radon} is the vectorial sum of the local standard of rest velocity, of the solar motion, and of the Earth's velocity with respect to the Sun. Here we consider a local standard of rest velocity of 220 km~s$^{-1}$, a solar motion of $(11.1, 12.2, 7.2)$ km~s$^{-1}$, and neglect the remaining (time-dependent) contribution to ${\bf v}_e(t)$. As for the Maxwell-Boltzmann approximation~(\ref{eq:MB}), this is a standard assumption regarding ${\bf v}_e(t)$, but it is not free from uncertainties. We refer to~\cite{Bovy:2012ba,Freese:2012xd,Strigari:2013iaa,McCabe:2013kea} for an exhaustive discussion on this subject. If not otherwise specified, we assume $\sigma_v=156$~km~s$^{-1}$, $\rho_\chi=0.4$~GeV~cm$^{-3}$, and $v_{\rm esc}=533$~km~s$^{-1}$~\cite{Catena:2009mf,Freese:2012xd,Piffl:2013mla}.
 
\section{Phenomenology}
\label{sec:pheno}

In this section we study the phenomenology of the effective theory of dark matter directional detection defined in Sec.~\ref{sec:theory}. The main observable we are interested in is the differential number of recoil events around $\cos\theta$ per unit time and per unit detector mass, namely ${\rm d}R/{\rm d}\hspace{-0.5mm}\cos\theta$ in Eq.~(\ref{eq:dR}). Using the Radon transforms of Sec.~\ref{sec:Radon}, and the nuclear response functions defined in Sec.~\ref{sec:nuc}, and given in Appendix~\ref{sec:appNuc}, we now compute ${\rm d}R/{\rm d}\hspace{-0.5mm}\cos\theta$ for all dark matter-nucleon interaction operators in Tab.~\ref{tab:operators}. In Secs.~\ref{sec:op1} and~\ref{sec:op2} we focus on scattering events induced by dark matter particles from the Milky Way dark matter halo. As an example of an alternative astrophysical configuration, in Sec.~\ref{sec:astro} we consider a hypothetical stream of dark matter particles moving in the direction of $-{\bf v}_e(t)$. If not otherwise specified, we assume $E_{\rm th}=0$.  Only in Sec.~\ref{sec:threshold} we relax this assumption, and quantify the impact of $E_{\rm th}\neq0$ on ${\rm d}R/{\rm d}\hspace{-0.5mm}\cos\theta$. 

In the following, the angle $\theta$ is measured with respect to the reference direction ${\bf v_e}(t)$. To simplify the comparison of different figures, in all plots we assume a single coupling constant different from zero and equal to $10^{-3}/m_v^2$ at the time, where $m_v=246.2$~GeV is the electroweak scale. Results are presented for three benchmark dark matter directional experiments  with head-tail discrimination made of CS$_2$, CF$_4$, and $^3$He, respectively. For definitiveness we consider spin-1/2 dark matter.

\subsection{Momentum/velocity independent operators}
\label{sec:op1}
We start with an analysis of the only momentum transfer and relative velocity independent operators, namely $\hat{\mathcal{O}}_1$ and $\hat{\mathcal{O}}_4$. In Eq.~(\ref{eq:Hx}), the operator $\hat{\mathcal{O}}_1$ contributes to the nuclear vector charge through the nuclear response operator $M_{LM;\tau}$, which in the low momentum transfer limit measures the nuclear mass number. The operator $\hat{\mathcal{O}}_4$ contributes to the nuclear spin current through $\Sigma'_{LM;\tau}$ and $\Sigma''_{LM;\tau}$, which in the low momentum transfer limit measure the nucleon spin content of the nucleus.

The cross-sections for dark matter-proton and dark matter-neutron scattering derived from the operator $\hat{\mathcal{O}}_1$ are
\begin{equation}
\label{eq:sigmaSI}
\sigma_{p}^{\rm SI} = \frac{\mu_{N}^2}{\pi} \, \frac{ |c_1^0 + c_1^1|^2}{4}\,; \qquad\qquad \sigma_{n}^{\rm SI} = \frac{\mu_{N}^2}{\pi} \, \frac{ |c_1^0 - c_1^1|^2}{4} \,,
\end{equation}
respectively. Similarly, from the operator $\hat{\mathcal{O}}_4$ one obtains the cross-sections for dark matter-proton and dark matter-neutron scattering
\begin{equation}
\label{eq:sigmaSD}
\sigma_{p}^{\rm SD} = \frac{\mu_N^2 j_\chi (j_\chi+1) }{4\pi}\, \frac{ |c_4^0 + c_4^1|^2}{4}\,; \qquad\qquad \sigma_{n}^{\rm SD} = \frac{\mu_N^2 j_\chi (j_\chi+1) }{4\pi}\, \frac{ |c_4^0 - c_4^1|^2}{4}  \,,
\end{equation}
respectively. 
Here $\mu_{N}=m_\chi m_{N}/(m_\chi+m_{N})$ is the reduced dark matter-nucleon mass, and $j_\chi=1/2$ is the dark matter particle spin. 

Eqs.~(\ref{eq:sigmaSI}) and~(\ref{eq:sigmaSD}) connect the formalism reviewed and developed here, to standard analyses of dark matter directional detection, where $c_1^0=c_1^1$ and $c_4^0=c_4^1$ is commonly assumed. In contrast, here we consider $c_1^0$, $c_1^1$, $c_4^0$, and $c_4^1$ as independent parameters, and compute ${\rm d}R/{\rm d}\hspace{-0.5mm}\cos\theta$ accordingly.

Fig.~\ref{fig:c1c4} shows the differential rate ${\rm d}R/{\rm d}\hspace{-0.5mm}\cos\theta$ as a function of $\cos\theta$ for the operators $\hat{\mathcal{O}}_1$ and $\hat{\mathcal{O}}_4$. In the left panels we focus on the isoscalar coupling constants $c_1^0$ and $c_4^0$, whereas the right panels are devoted to the isovector coupling constants $c_1^1$ and $c_4^1$, respectively. In the plots different colors and lines refer to the target materials, and to the values of the dark matter particle mass reported in the legends. 

The isoscalar component of the operator $\hat{\mathcal{O}}_1$ can be probed by detectors made of CS$_2$, CF$_4$, and $^{3}$He. In contrast, a detector composed of CS$_2$ would be insensitive to the isovector component of the operator $\hat{\mathcal{O}}_1$. 

Regarding the operator $\hat{\mathcal{O}}_4$, we find that this interaction cannot be detected using CS$_2$ as a target material, as both $^{12}$C and $^{32}$S have spin zero. We also observe that the isoscalar and isovector components of the operator $\hat{\mathcal{O}}_4$ generate a similar differential rate ${\rm d}R/{\rm d}\hspace{-0.5mm}\cos\theta$, as the nuclear response functions $W^{00}_{\Sigma'}$,  $W^{11}_{\Sigma'}$, $W^{00}_{\Sigma''}$,  and $W^{11}_{\Sigma''}$ differ by about a factor of 2 for $^{19}$F and $^{3}$He (see appendix~\ref{sec:appNuc}).

\subsection{Momentum/velocity dependent operators}
\label{sec:op2}
Next, we focus on the dark matter-nucleon interactions in Tab.~\ref{tab:operators} depending on the momentum transfer operator ${\bf\hat{q}}$, on the dark matter-nucleus transverse relative velocity operator ${\bf\hat{v}}^{\perp}$, or on both.

Contrary to $\hat{\mathcal{O}}_1$ and $\hat{\mathcal{O}}_4$, for these operators the differential rate ${\rm d}R/{\rm d}\hspace{-0.5mm}\cos\theta$ can depend on nuclear response functions that do not appear in the theory of electroweak scattering from nuclei, and are specific to one-body dark matter-nucleon interactions. An example of nuclear response functions that are not generated by the operators $\hat{\mathcal{O}}_1$ and $\hat{\mathcal{O}}_4$ is $W^{\tau\tau'}_{\Phi''}$, which is different from zero for $^{12}$C, $^{19}$F and $^{32}$S. The corresponding nuclear response operator $\Phi''_{LM;\tau}(q )$ measures the content of nucleon spin-orbit coupling in the nucleus. Therefore, $W^{\tau\tau'}_{\Phi''}$ can be large for isotopes with non fully occupied single-nucleon orbits of large angular momentum~\cite{Fitzpatrick:2012ix}.

Figs.~\ref{fig:c3c5}, \ref{fig:c6c7}, \ref{fig:c8c9}, \ref{fig:c10c11}, \ref{fig:c12c13} and \ref{fig:c14c15} show the differential rate ${\rm d}R/{\rm d}\hspace{-0.5mm}\cos\theta$ as a function of $\cos\theta$ for the interaction operators in Tab.~\ref{tab:operators} that depend on ${\bf\hat{q}}$ and/or ${\bf\hat{v}}^{\perp}$. Different colors correspond to distinct target materials. As in the previous section, we compute ${\rm d}R/{\rm d}\hspace{-0.5mm}\cos\theta$ for directional detectors with head-tail discrimination made of CS$_2$, CF$_4$ and $^3$He. Solid, dashed and dot-dashed lines denote a dark matter particle mass $m_\chi$ of 10 GeV, 100 GeV and 1000 GeV, respectively. 

The amplitude of the differential rate ${\rm d}R/{\rm d}\hspace{-0.5mm}\cos\theta$ for a given interaction operator $\hat{\mathcal{O}}_k$ reflects the number of ${\bf\hat{q}}$ and ${\bf\hat{v}}^{\perp}$ operators that multiply $c_k^\tau$ in Eq.~(\ref{eq:ls}). However, it also crucially depends on the nuclear response functions in Appendix~\ref{sec:appNuc} integrated over energies larger than $E_{\rm th}$.

As an example of the importance that nuclear response functions can have in the calculation of ${\rm d}R/{\rm d}\hspace{-0.5mm}\cos\theta$, let us focus on the operators $\hat{\mathcal{O}}_{4}={\bf{\hat{S}}}_{\chi}\cdot {\bf{\hat{S}}}_{N}$ and $\hat{\mathcal{O}}_{11}=i{\bf{\hat{S}}}_\chi\cdot{\bf{\hat{q}}}/m_N$. The former generates a nuclear spin current, the latter a nuclear vector charge. We find that for CF$_4$ detectors, the isoscalar component of the operator $\hat{\mathcal{O}}_{11}$ contributes to ${\rm d}R/{\rm d}\hspace{-0.5mm}\cos\theta$ with the same strength of the isoscalar component of the operator $\hat{\mathcal{O}}_{4}$, though $\hat{\mathcal{O}}_{4}$ is momentum independent. For $^{12}$C and $^{19}$ F, the response function $W^{00}_{M}$ is larger than $W^{00}_{\Sigma'}$ and $W^{00}_{\Sigma''}$, and it compensates for the momentum suppression characterizing the operator $\hat{\mathcal{O}}_{11}$.
Interestingly, in the literature the operator $\hat{\mathcal{O}}_{11}$ is much less explored than the familiar spin-dependent operator $\hat{\mathcal{O}}_4$.

In Figs.~\ref{fig:c3c5}, \ref{fig:c6c7}, \ref{fig:c8c9}, \ref{fig:c10c11}, \ref{fig:c12c13} and \ref{fig:c14c15}, we also observe that (besides 2 exceptions discussed below) the spin-dependent operators in Tab.~\ref{tab:operators}, i.e. ${\bf{\hat{S}}}_{N}$ dependent, do not produce any signal at CS$_2$ detectors, as both $^{12}$C and $^{32}$S have spin 0. The operators $\hat{\mathcal{O}}_{12} = {\bf{\hat{S}}}_{\chi}\cdot ({\bf{\hat{S}}}_{N} \times{\bf{\hat{v}}}^{\perp})$ and $\hat{\mathcal{O}}_{15} = -({\bf{\hat{S}}}_{\chi}\cdot {\bf{\hat{q}}}/m_N)[ ({\bf{\hat{S}}}_{N}\times {\bf{\hat{v}}}^{\perp}) \cdot {\bf{\hat{q}}}/m_N] $ constitute 2 interesting exceptions. Though ${\bf{\hat{S}}}_{N}$ dependent, they do not only induce a dark matter coupling to the nuclear spin. They also generate the nuclear response function $W_{\Phi''}^{\tau\tau'}$, which is different from zero for $^{12}$C and $^{32}$S. Interestingly, detectors made of CS$_2$ cannot probe any dark matter-nucleon isovector coupling. 

\begin{figure}[t]
\begin{center}
\begin{minipage}[t]{0.49\linewidth}
\centering
\includegraphics[width=\textwidth]{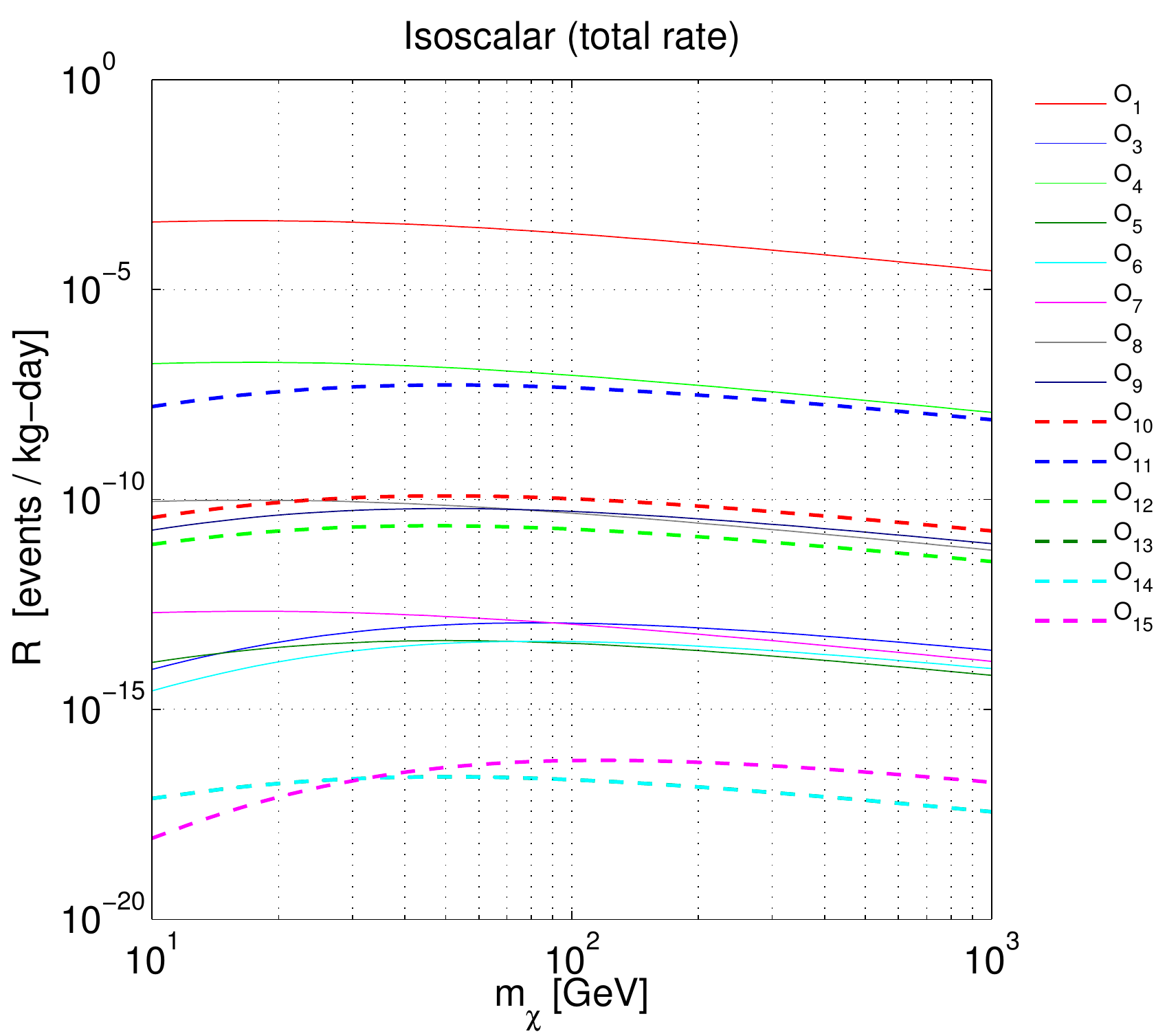}
\end{minipage}
\begin{minipage}[t]{0.49\linewidth}
\centering
\includegraphics[width=\textwidth]{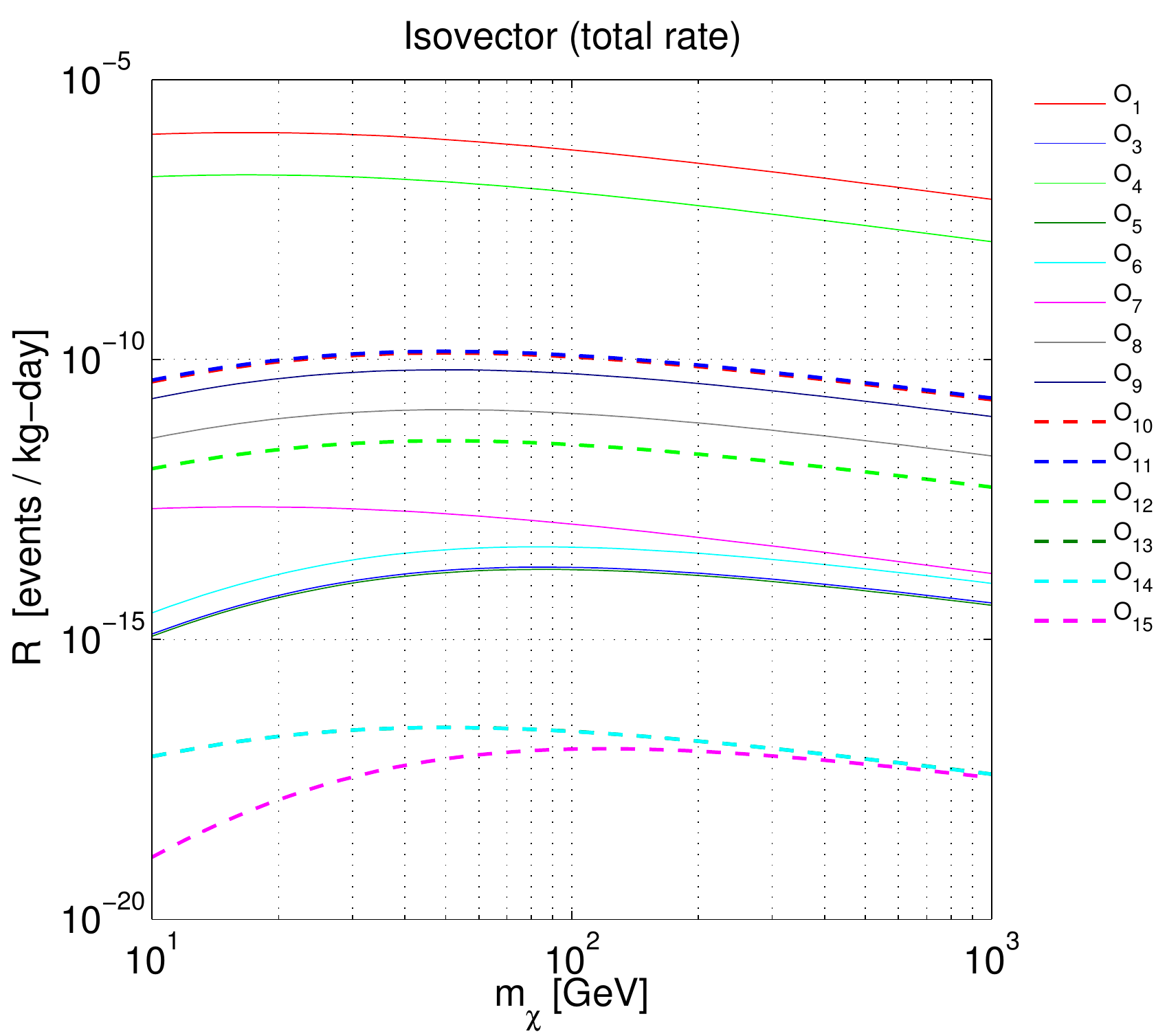}
\end{minipage}
\end{center}
\caption{Total rate $R$ as a function of $m_\chi$ for all operators in the legends. The left panel refers to the isoscalar couplings, whereas the right panel to the isovector couplings. For operators  with differential cross-sections~(\ref{eq:sigma}) independent of $q$, or depending on $q$ trough $v_T^{\perp 2}$ only, the total rate $R$ grows with $m_\chi$ up to $m_\chi\sim10$~GeV, and then it decreases for larger masses. For operators with differential cross-sections explicitly depending on $q$, $R$ grows up to $m_\chi$ of the order of 100 GeV, and then it decreases for larger values of $m_\chi$. }
\label{fig:Rate}
\end{figure}

We conclude this section with a brief discussion on the $m_\chi$ dependence of the differential rate~(\ref{eq:dR}). The rate ${\rm d}R/{\rm d}\hspace{-0.5mm}\cos\theta$ depends on $m_\chi$ through the minimum velocity $w_T$, and the dark matter number density $(\rho_\chi/m_\chi)$. The factor $(\rho_\chi/m_\chi)$ in Eq.~(\ref{eq:dR}) is the same for all operators in Tab.~\ref{tab:operators}. 
The dependence on $m_\chi$ of ${\rm d}R/{\rm d}\hspace{-0.5mm}\cos\theta$ is therefore non trivially determined by the operator-dependent double integral in Eq.~(\ref{eq:dR}). 
For a given interaction operator, the dependence of ${\rm d}R/{\rm d}\hspace{-0.5mm}\cos\theta$ on $m_\chi$ can be qualitatively understood from a plot of the total scattering rate as function of $m_\chi$. Fig.~\ref{fig:Rate} shows the total rate $R$, defined as
\begin{equation}
R = \int_{-1}^{+1} {\rm d}\hspace{-0.5mm}\cos\theta \, \frac{{\rm d}R}{{\rm d}\hspace{-0.5mm}\cos\theta}\,,
\end{equation}
as a function of $m_\chi$ for all operators in Tab.~\ref{tab:operators}. We find that for operators with differential cross-sections~(\ref{eq:sigma}) independent of $q$, or depending on $q$ trough $v_T^{\perp 2}$ only, the total rate $R$ grows with $m_\chi$ up to $m_\chi\sim10$~GeV, and then it decreases for larger masses. This applies to the operators $\hat{\mathcal{O}}_1$, $\hat{\mathcal{O}}_4$, $\hat{\mathcal{O}}_7$, and $\hat{\mathcal{O}}_8$. For operators with differential cross-sections explicitly depending on momentum, i.e. not through $v_{T}^{\perp}$ only, $R$ grows up to $m_\chi$ of the order of 100 GeV, and then it decreases for larger values of $m_\chi$. The $m_\chi$ dependence observed in Fig.~\ref{fig:Rate} reflects in Figs.~\ref{fig:c3c5}, \ref{fig:c6c7}, \ref{fig:c8c9}, \ref{fig:c10c11}, \ref{fig:c12c13} and \ref{fig:c14c15} accordingly.

\subsection{Changing astrophysical assumptions}
\label{sec:astro}
In the limit $\sigma_v/|{\bf v}_e|\rightarrow 0$, the equations derived in Sec.~\ref{sec:astro} can also be applied to the directional detection of dark matter particles from galactic streams. Here we focus on a hypothetical stream of dark matter particles moving at a speed of 300~km~s$^{-1}$ in the direction of $-{\bf v}_e$, with velocity dispersion $\sigma_v=5$~km~s$^{-1}$ and $\rho_\chi=0.3$~GeV~cm$^{-3}$.

\begin{figure}[t]
\begin{center}
\begin{minipage}[t]{0.49\linewidth}
\centering
\includegraphics[width=\textwidth]{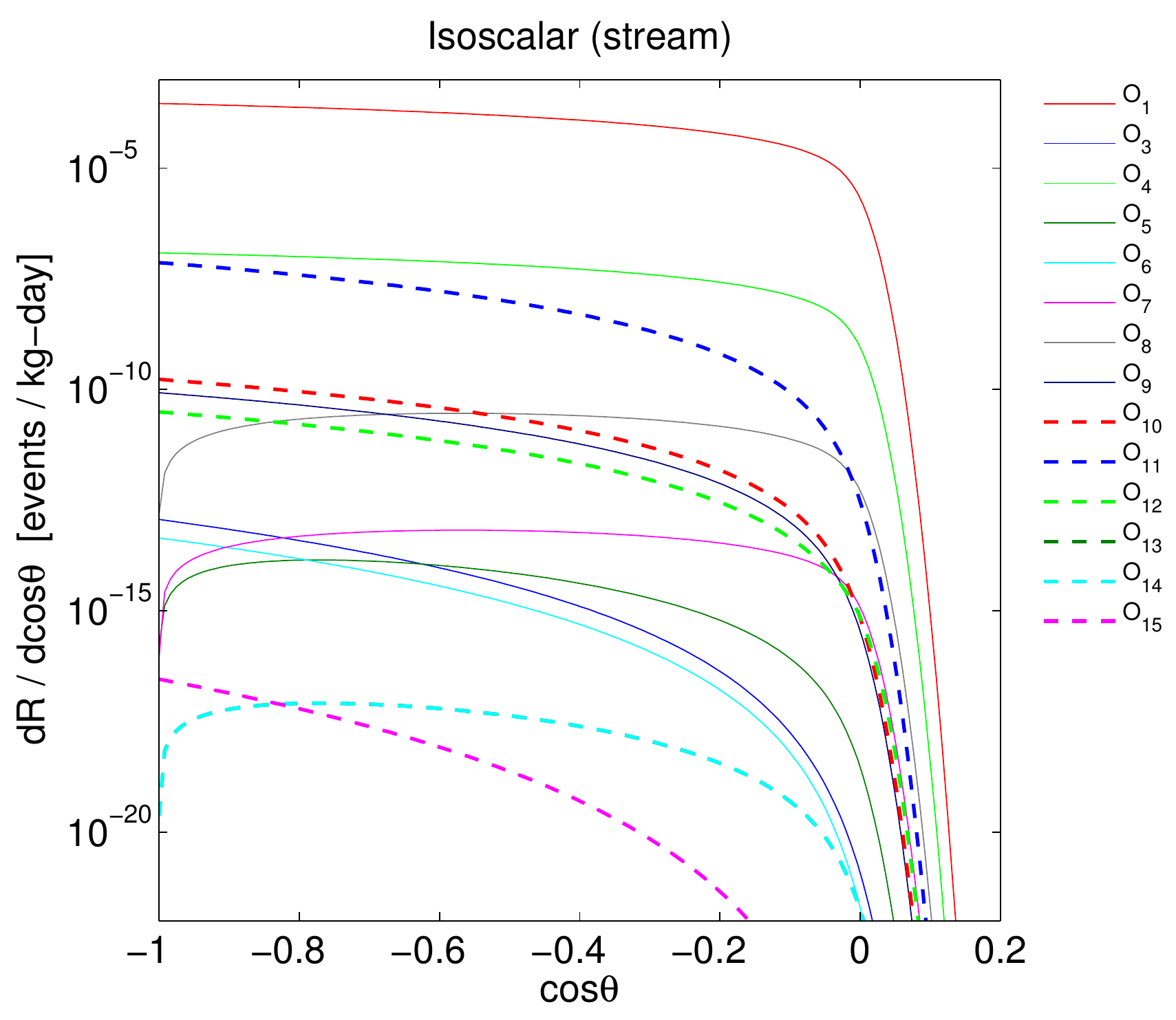}
\end{minipage}
\begin{minipage}[t]{0.49\linewidth}
\centering
\includegraphics[width=\textwidth]{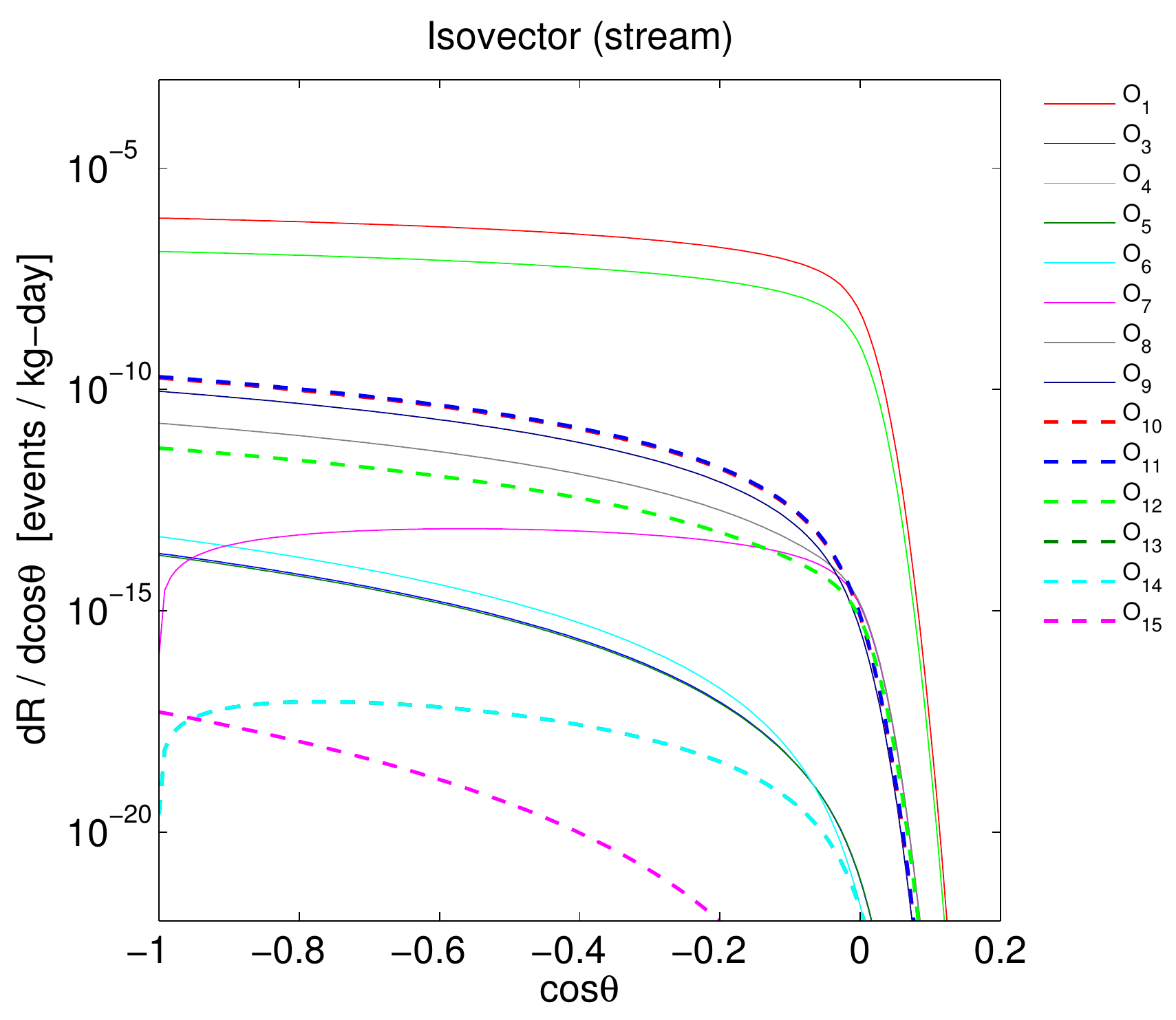}
\end{minipage}
\end{center}
\caption{Differential number of recoil events around $\cos\theta$ per unit time and per unit detector mass as a function of $\cos\theta$ for a stream of dark matter particles moving at a speed of 300~km~s$^{-1}$ in the direction of $-{\bf v}_e$, with velocity dispersion $\sigma_v=5$~km~s$^{-1}$, and $\rho_\chi=0.3$~GeV~cm$^{-3}$. We report ${\rm d}R/{\rm d}\hspace{-0.5mm}\cos\theta$ for a directional detector with head-tail discrimination made of CF$_4$, and assume $m_\chi=100$~GeV. The left panel and the right panel refer, respectively, to the isoscalar and isovector components of the operators in Tab.~\ref{tab:operators}, as shown in the legends.}
\label{fig:Stream}
\end{figure}

Fig.~\ref{fig:Stream} shows the differential rate ${\rm d}R/{\rm d}\hspace{-0.5mm}\cos\theta$ as a function of $\cos\theta$ for the hypothetical dark matter stream described above, and for all dark matter-nucleon interaction operators in Tab.~\ref{tab:operators}. Here we consider CF$_4$ as a target material, $m_\chi=100$~GeV as a benchmark dark matter particle mass, and $E_{\rm th}=0$.

For interaction operators independent of ${\bf{v}}^{\perp}$, the differential rate ${\rm d}R/{\rm d}\hspace{-0.5mm}\cos\theta$ has a maximum at $\cos\theta=-1$, and it rapidly goes to zero around $\cos\theta=0$, as expected for the stream considered here.

For interaction operators that do depend on ${\bf{v}}^{\perp}$, ${\rm d}R/{\rm d}\hspace{-0.5mm}\cos\theta$ is linear in $\hat{f}^{(2)}_M(w_T,{\bf w})$, and it therefore exhibits a different behavior. For these operators ${\rm d}R/{\rm d}\hspace{-0.5mm}\cos\theta$ has a maximum at larger values of $\cos\theta$, because of the $\hat{f}^{(2)}_M(w_T,{\bf w})$ contribution to the differential rate. This interesting effect is not present when only the interaction operators $\hat{\mathcal{O}}_1$ and  $\hat{\mathcal{O}}_4$ are considered. The shift in the peak of ${\rm d}R/{\rm d}\hspace{-0.5mm}\cos\theta$ is pronounced for the interaction operators $\hat{\mathcal{O}}_5$, $\hat{\mathcal{O}}_7$, $\hat{\mathcal{O}}_8$, $\hat{\mathcal{O}}_{13}$, and $\hat{\mathcal{O}}_{14}$. For the ${\bf{v}}^{\perp}$ dependent operators $\hat{\mathcal{O}}_{3}$, $\hat{\mathcal{O}}_{12}$ and $\hat{\mathcal{O}}_{15}$ the effect is instead negligible, as for these operators the contribution of $\hat{f}^{(2)}_M(w_T,{\bf w})$ to ${\rm d}R/{\rm d}\hspace{-0.5mm}\cos\theta$ is sub-leading (see also Eq.~(\ref{eq:R}) in Appendix~\ref{sec:appDM}).

\begin{figure}[t]
\begin{center}
\begin{minipage}[t]{0.49\linewidth}
\centering
\includegraphics[width=\textwidth]{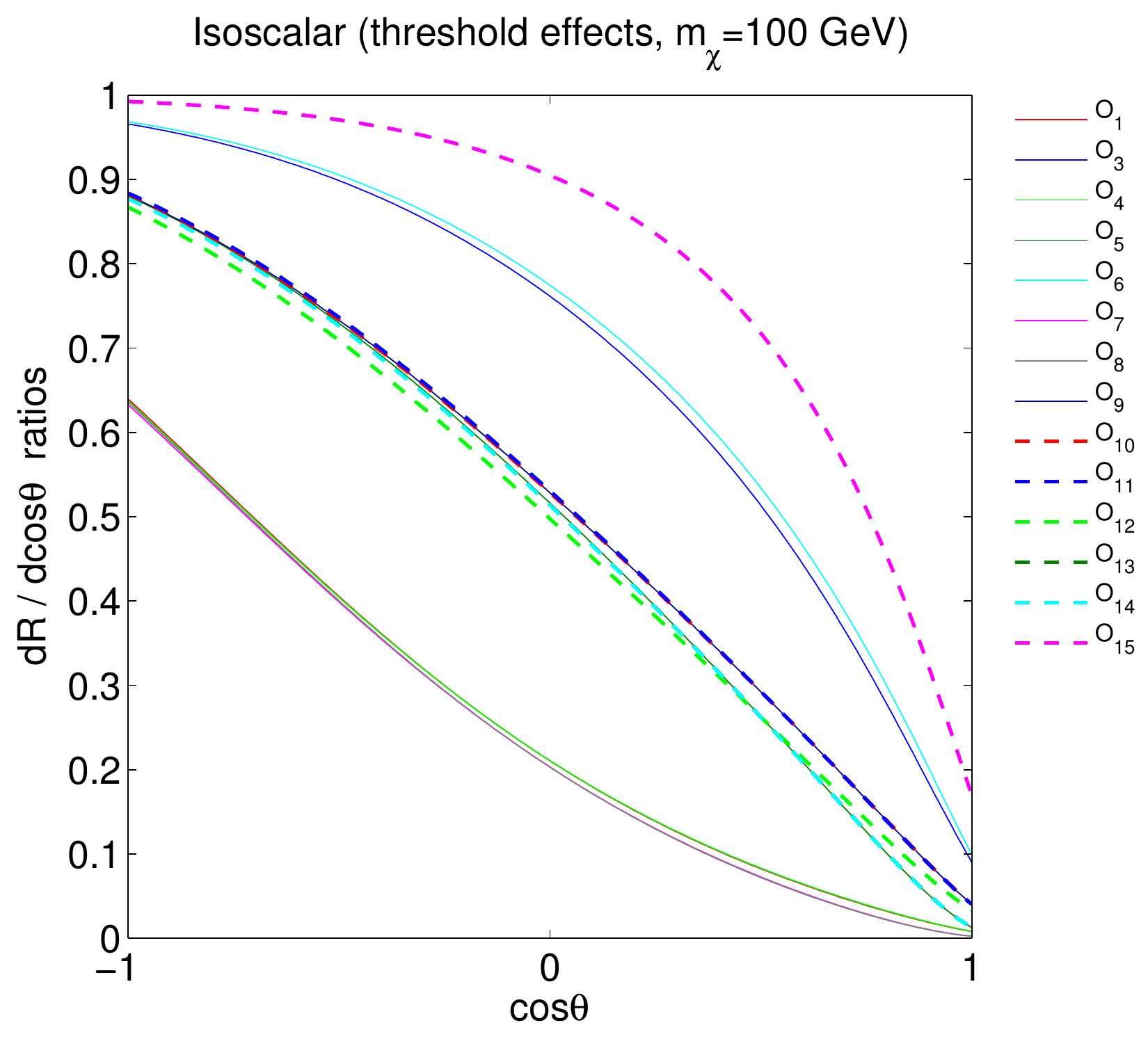}
\end{minipage}
\begin{minipage}[t]{0.49\linewidth}
\centering
\includegraphics[width=\textwidth]{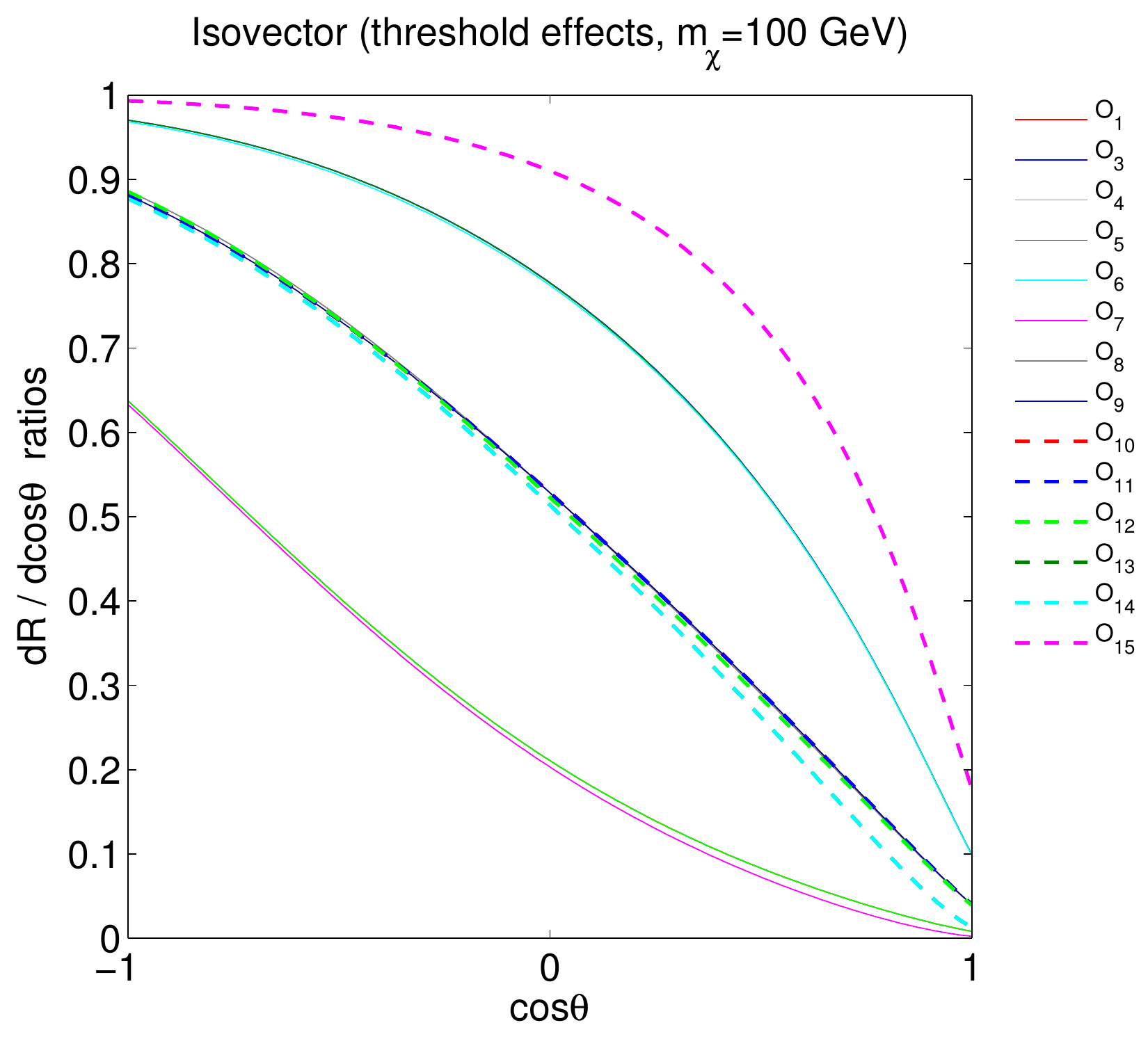}
\end{minipage}
\begin{minipage}[t]{0.49\linewidth}
\centering
\includegraphics[width=\textwidth]{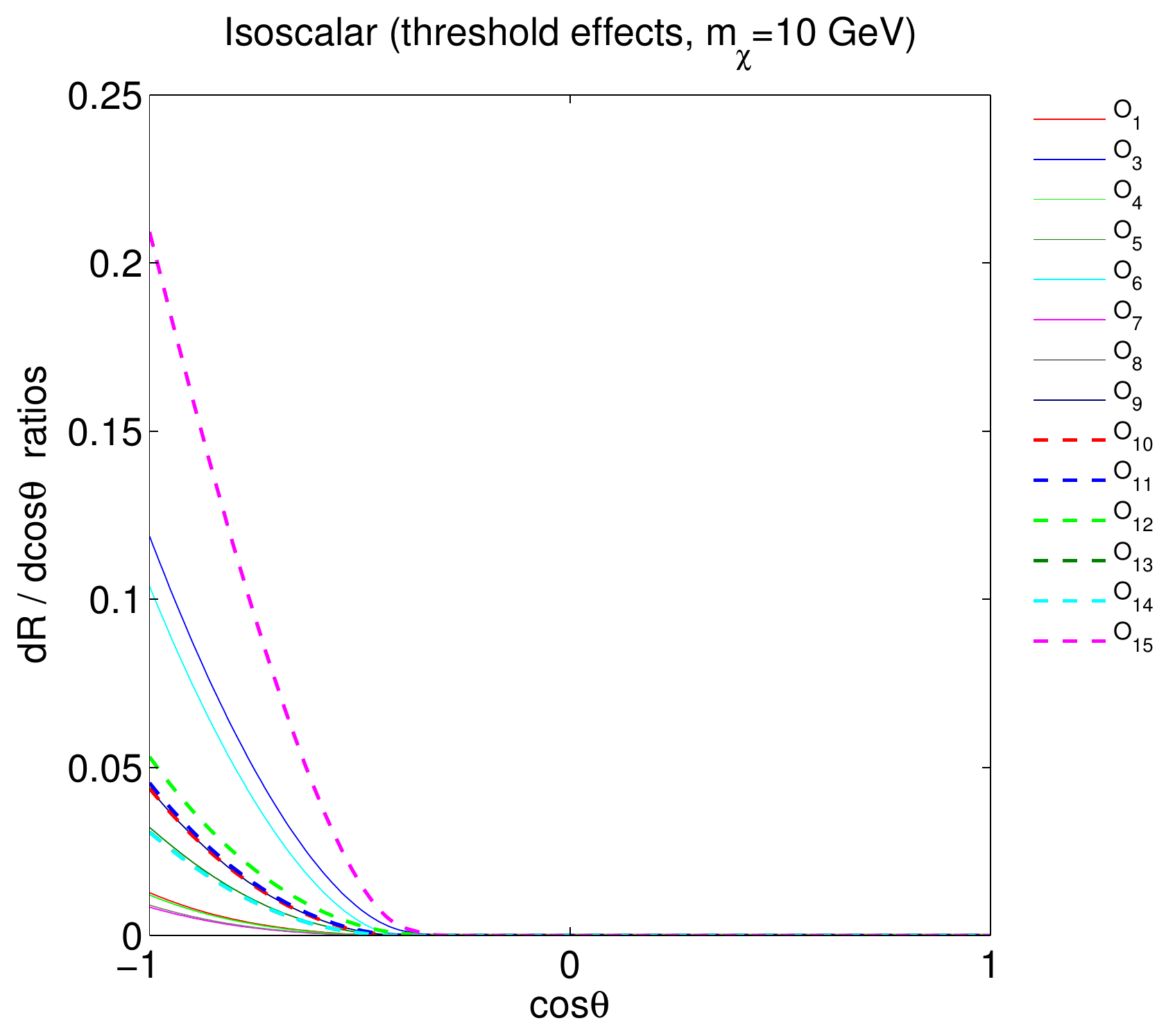}
\end{minipage}
\begin{minipage}[t]{0.49\linewidth}
\centering
\includegraphics[width=\textwidth]{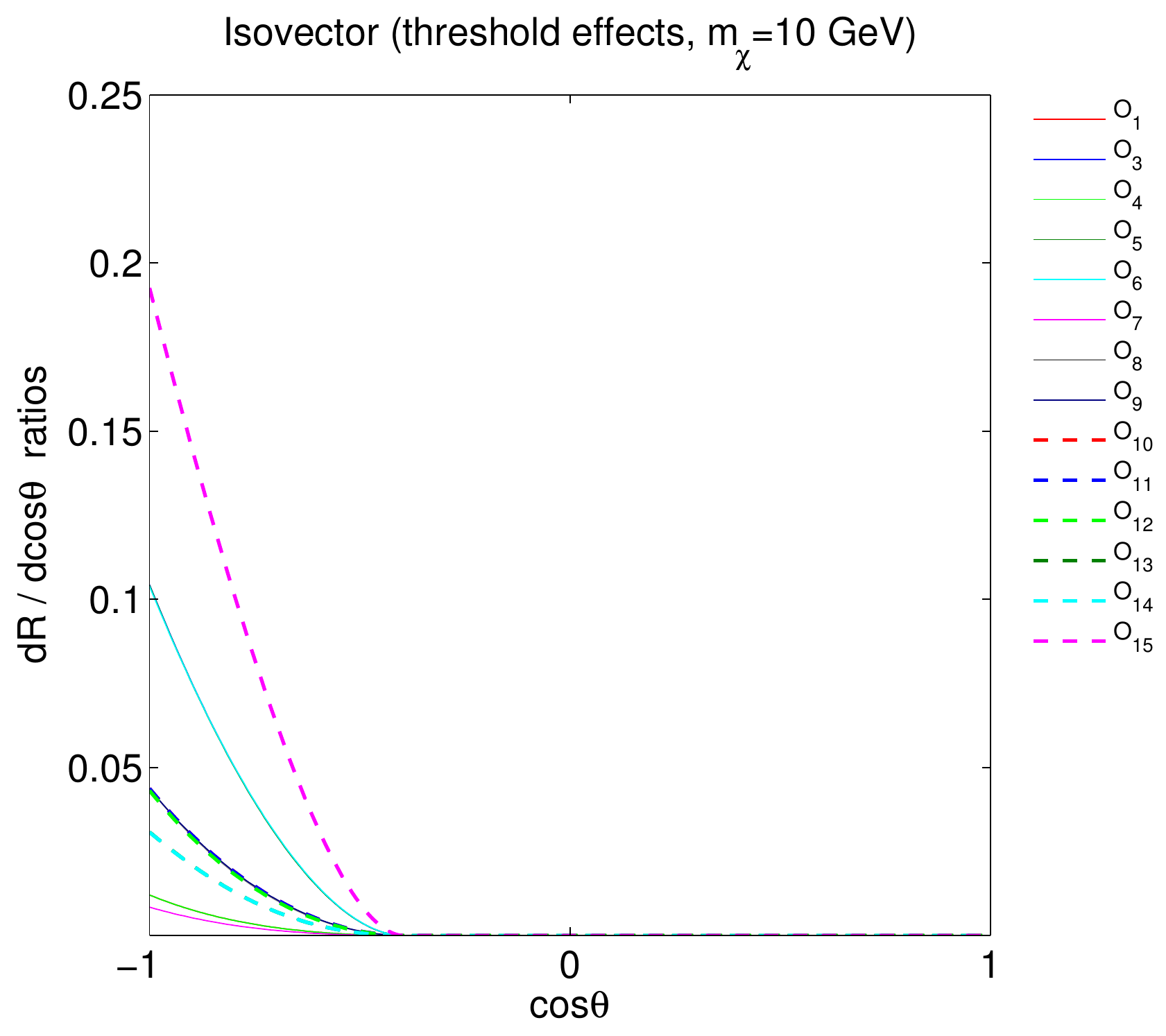}
\end{minipage}
\end{center}
\caption{Ratio of the differential rate ${\rm d}R/{\rm d}\hspace{-0.5mm}\cos\theta$ for $E_{\rm th}=20$~keV to the same rate with $E_{\rm th}=0$, for all interaction operators in Tab.~\ref{tab:operators}. Colors and lines refer to the operators reported in the legends. The top panels correspond to $m_\chi=100$~GeV, whereas the bottom panels refer to $m_\chi=10$~GeV.}
\label{fig:Th}
\end{figure}
\subsection{Threshold effects}
\label{sec:threshold}
We now relax the assumption $E_{\rm th}=0$, and study the impact of a finite experimental energy threshold on the differential rate ${\rm d}R/{\rm d}\hspace{-0.5mm}\cos\theta$. Here we consider $E_{\rm th}=20$~keV, though lower energy thresholds are expected for the future~\cite{Ahlen:2009ev}. We restrict the analysis to a directional detector made of CF$_{4}$, and consider 10 GeV and 100 GeV as benchmark dark matter particle masses.

Fig.~\ref{fig:Th} shows the ratio of the differential rate ${\rm d}R/{\rm d}\hspace{-0.5mm}\cos\theta$ for $E_{\rm th}=20$~keV to the same rate with $E_{\rm th}=0$ for all interaction operators in Tab.~\ref{tab:operators}. Colors and lines refer to the operators reported in the legends. 

Regarding threshold effects, we find that the operators in Tab.~\ref{tab:operators} divide into 4 independent subsets. A first subset includes the operators $\hat{\mathcal{O}}_1$, $\hat{\mathcal{O}}_4$, $\hat{\mathcal{O}}_7$, and $\hat{\mathcal{O}}_8$. These are the operators most sensitive to a finite energy threshold. They contribute to the differential rate~(\ref{eq:dR}) through dark matter response functions either constant, or proportional to $v_T^2$. The remaining 3 subsets of operators are increasingly less sensitive to threshold effects. The second subset of operators involves $\hat{\mathcal{O}}_{5}$, $\hat{\mathcal{O}}_{9}$, $\hat{\mathcal{O}}_{10}$, $\hat{\mathcal{O}}_{11}$, $\hat{\mathcal{O}}_{12}$, $\hat{\mathcal{O}}_{13}$, and $\hat{\mathcal{O}}_{14}$. The dark matter response functions of these operators are either proportional to $q^2$, or to $q^2\,v_T^2$ ($v_T^2$ contributions are sub-leading). The third and fourth subsets consist of $\hat{\mathcal{O}}_{3}$ and $\hat{\mathcal{O}}_{6}$, and of $\hat{\mathcal{O}}_{15}$ alone, respectively. The former is characterized by dark matter response functions 
proportional to $q^4$ ($q^2\,v_T^2$ contributions are sub-leading). The latter by response functions proportional to $q^6$ ($q^4\,v_T^2$ contributions are sub-leading).

The dependence on the momentum transfer specific to operators in different subsets reflects in the 4 distinct behaviors that we observe in the top panels ($m_\chi=100$~GeV) and in the bottom panels ($m_\chi=10$~GeV) of Fig.~\ref{fig:Th}. As expected, for low dark matter particle masses, and for $\cos\theta \gtrsim-1$, threshold effects are more pronounced. 

\section{Conclusions}
\label{sec:conc}

We have extended the formalism of dark matter directional detection to arbitrary one-body dark matter-nucleon interactions. The resulting theoretical framework predicts 28 independent recoil energy spectra characterized by 8 isotope-dependent nuclear response functions, and by the Radon transforms of the first and second moment of the local dark matter velocity distribution. In this study, we have analytically computed the Radon transform of the first 2 moments of a truncated Maxwell-Boltzmann distribution, and made use of nuclear response functions recently obtained in the literature through numerical nuclear structure calculations. 

We have computed the rate ${\rm d}R/{\rm d}\hspace{-0.5mm}\cos\theta$ as a function of the nuclear recoil angle 
for the 28 isoscalar and isovector dark matter-nucleon interactions of the theory. We have presented our results for 3 hypothetical dark matter directional detectors with head-tail discrimination composed of CS$_2$, CF$_4$ and $^3$He, respectively. The relevant nuclear response functions for the isotopes $^{3}$He, $^{12}$C, $^{19}$F and $^{32}$S are listed in Appendix~\ref{sec:appNuc}. We have separately considered the case of dark matter particles from the Milky Way dark matter halo, and of a dark matter stream.

We have found that only CF$_4$ and $^3$He detectors can probe all interactions explored in this work ($^3$He detectors with a typically smaller rate). CS$_2$ detectors are insensitive to dark matter-nucleon interaction operators that couple to the nuclear spin-current only, as both $^{12}$C and $^{32}$S have zero spin. Interestingly, we have found that the spin-dependent operators $\hat{\mathcal{O}}_{12} = {\bf{\hat{S}}}_{\chi}\cdot ({\bf{\hat{S}}}_{N} \times{\bf{\hat{v}}}^{\perp})$ and $\hat{\mathcal{O}}_{15} = -({\bf{\hat{S}}}_{\chi}\cdot {\bf{\hat{q}}}/m_N)[ ({\bf{\hat{S}}}_{N}\times {\bf{\hat{v}}}^{\perp}) \cdot {\bf{\hat{q}}}/m_N] $ remain accessible to CS$_2$ detectors, in that the operators $\hat{\mathcal{O}}_{12}$ and $\hat{\mathcal{O}}_{15}$ generate a non zero nuclear spin-velocity current.

We have also found that for CF$_4$ detectors the operator $\hat{\mathcal{O}}_{11}=i{\bf{\hat{S}}}_\chi\cdot{\bf{\hat{q}}}/m_N$ can contribute to ${\rm d}R/{\rm d}\hspace{-0.5mm}\cos\theta$ with the same strength of the more popular operator $\hat{\mathcal{O}}_{4}={\bf{\hat{S}}}_{\chi}\cdot {\bf{\hat{S}}}_{N}$, though $\hat{\mathcal{O}}_{4}$ is momentum independent.~This intriguing result highlights the importance of nuclear response functions, and numerical nuclear structure calculations in dark matter directional detection. 

We have characterized the dependence on $m_\chi$, and on the experimental energy threshold $E_{\rm th}$ of the rate ${\rm d}R/{\rm d}\hspace{-0.5mm}\cos\theta$ for all one-body dark matter-nucleon interactions in Tab.~\ref{tab:operators}. For a given dark matter-nucleon interaction, both properties reflect the momentum transfer dependence of the differential scattering cross-section. We have found that the stronger the differential cross-section depends on the momentum transfer the less a specific interaction is sensitive to $E_{\rm th}$.

Finally, we have found that the rate ${\rm d}R/{\rm d}\hspace{-0.5mm}\cos\theta$ for the operators $\hat{\mathcal{O}}_5$, $\hat{\mathcal{O}}_7$, $\hat{\mathcal{O}}_8$, $\hat{\mathcal{O}}_{13}$, and $\hat{\mathcal{O}}_{14}$ has not a maximum in the direction opposite to the observer's motion. The shift in the peak of ${\rm d}R/{\rm d}\hspace{-0.5mm}\cos\theta$ is pronounced for dark matter streams.~This effect is present whenever the second moment of the dark matter velocity distribution is quantitatively important, as for the operators listed above. This feature is similar to the rings found in~\cite{Bozorgnia:2011vc} studying ${\rm d}^2R/{\rm d}\hspace{-0.5mm}\cos\theta{\rm d}E_R$, but it characterizes the rate ${\rm d}R/{\rm d}\hspace{-0.5mm}\cos\theta$, which is integrated above the energy threshold $E_{\rm th}$. 

Our study shows that the phenomenology of dark matter directional detection can be more complex and rich than in current analyses based on the familiar operators $\hat{\mathcal{O}}_1$ and $\hat{\mathcal{O}}_4$ only. The equations reported in this work can be used for model independent analyses of dark matter signals at directional detection experiments.

\acknowledgments This work has partially been funded through a start-up grant of the University of G\"ottingen. R.C. acknowledges partial support from the European Union FP7 ITN INVISIBLES (Marie Curie Actions, PITN-GA-2011-289442).

\appendix

\section{Dark matter response functions}
\label{sec:appDM}
Below, we list the dark matter response functions that appear in Eq.~(\ref{eq:sigma}). The notation is the same used in the body of the paper.
\begin{eqnarray}
 R_{M}^{\tau \tau^\prime}\left(v_T^{\perp 2}, {q^2 \over m_N^2}\right) &=& c_1^\tau c_1^{\tau^\prime } + {j_\chi (j_\chi+1) \over 3} \left[ {q^2 \over m_N^2} v_T^{\perp 2} c_5^\tau c_5^{\tau^\prime }+v_T^{\perp 2}c_8^\tau c_8^{\tau^\prime }
+ {q^2 \over m_N^2} c_{11}^\tau c_{11}^{\tau^\prime } \right] \nonumber \\
 R_{\Phi^{\prime \prime}}^{\tau \tau^\prime}\left(v_T^{\perp 2}, {q^2 \over m_N^2}\right) &=& {q^2 \over 4 m_N^2} c_3^\tau c_3^{\tau^\prime } + {j_\chi (j_\chi+1) \over 12} \left( c_{12}^\tau-{q^2 \over m_N^2} c_{15}^\tau\right) \left( c_{12}^{\tau^\prime }-{q^2 \over m_N^2}c_{15}^{\tau^\prime} \right)  \nonumber \\
 R_{\Phi^{\prime \prime} M}^{\tau \tau^\prime}\left(v_T^{\perp 2}, {q^2 \over m_N^2}\right) &=&  c_3^\tau c_1^{\tau^\prime } + {j_\chi (j_\chi+1) \over 3} \left( c_{12}^\tau -{q^2 \over m_N^2} c_{15}^\tau \right) c_{11}^{\tau^\prime } \nonumber \\
  R_{\tilde{\Phi}^\prime}^{\tau \tau^\prime}\left(v_T^{\perp 2}, {q^2 \over m_N^2}\right) &=&{j_\chi (j_\chi+1) \over 12} \left[ c_{12}^\tau c_{12}^{\tau^\prime }+{q^2 \over m_N^2}  c_{13}^\tau c_{13}^{\tau^\prime}  \right] \nonumber \\
   R_{\Sigma^{\prime \prime}}^{\tau \tau^\prime}\left(v_T^{\perp 2}, {q^2 \over m_N^2}\right)  &=&{q^2 \over 4 m_N^2} c_{10}^\tau  c_{10}^{\tau^\prime } +
  {j_\chi (j_\chi+1) \over 12} \left[ c_4^\tau c_4^{\tau^\prime} + \right.  \nonumber \\
 && \left. {q^2 \over m_N^2} ( c_4^\tau c_6^{\tau^\prime }+c_6^\tau c_4^{\tau^\prime })+
 {q^4 \over m_N^4} c_{6}^\tau c_{6}^{\tau^\prime } +v_T^{\perp 2} c_{12}^\tau c_{12}^{\tau^\prime }+{q^2 \over m_N^2} v_T^{\perp 2} c_{13}^\tau c_{13}^{\tau^\prime } \right] \nonumber \\
    R_{\Sigma^\prime}^{\tau \tau^\prime}\left(v_T^{\perp 2}, {q^2 \over m_N^2}\right)  &=&{1 \over 8} \left[ {q^2 \over  m_N^2}  v_T^{\perp 2} c_{3}^\tau  c_{3}^{\tau^\prime } + v_T^{\perp 2}  c_{7}^\tau  c_{7}^{\tau^\prime }  \right]
       + {j_\chi (j_\chi+1) \over 12} \left[ c_4^\tau c_4^{\tau^\prime} +  \right.\nonumber \\
       &&\left. {q^2 \over m_N^2} c_9^\tau c_9^{\tau^\prime }+{v_T^{\perp 2} \over 2} \left(c_{12}^\tau-{q^2 \over m_N^2}c_{15}^\tau \right) \left( c_{12}^{\tau^\prime }-{q^2 \over m_N^2}c_{15}^{\tau \prime} \right) +{q^2 \over 2 m_N^2} v_T^{\perp 2}  c_{14}^\tau c_{14}^{\tau^\prime } \right] \nonumber \\
     R_{\Delta}^{\tau \tau^\prime}\left(v_T^{\perp 2}, {q^2 \over m_N^2}\right)&=&  {j_\chi (j_\chi+1) \over 3} \left[ {q^2 \over m_N^2} c_{5}^\tau c_{5}^{\tau^\prime }+ c_{8}^\tau c_{8}^{\tau^\prime } \right] \nonumber \\
 R_{\Delta \Sigma^\prime}^{\tau \tau^\prime}\left(v_T^{\perp 2}, {q^2 \over m_N^2}\right)&=& {j_\chi (j_\chi+1) \over 3} \left[c_{5}^\tau c_{4}^{\tau^\prime }-c_8^\tau c_9^{\tau^\prime} \right].
 \label{eq:R}
\end{eqnarray}

\section{Nuclear response functions}
\label{sec:appNuc}
Below, we list the nuclear response functions relevant for dark matter directional detection and different from zero. For $^{3}$He, $^{12}$C, and $^{32}$S, we use the nuclear response functions that we obtained in~\cite{Catena:2015uha}. For $^{19}$F we adopt the nuclear response functions found in~\cite{Fitzpatrick:2012ix}.

\subsection*{Helium ($^3$He)}
\begin{flalign}
W^{00}_{M}(y)&= 0.358099 e^{-2y}& W^{00}_{\Sigma^{\prime\prime}}(y)&= 0.0397887  e^{-2y} &W^{00}_{\Sigma^\prime}(y)&= 0.0795775  e^{-2y} &\nonumber\\
W^{11}_{M}(y)&= 0.0397887 e^{-2y}& W^{11}_{\Sigma^{\prime\prime}}(y)&= 0.0397887 e^{-2y} &W^{11}_{\Sigma^\prime}(y)&= 0.0795775 e^{-2y} &  \nonumber\\ 
W^{10}_{M}(y)&= 0.119366 e^{-2y} &W^{10}_{\Sigma^{\prime\prime}}(y)&= -0.0397887 e^{-2y} &W^{10}_{\Sigma^\prime}(y)&= -0.0795775 e^{-2y} &  \nonumber\\ 
W^{01}_{M}(y)&= 0.119366 e^{-2y}& W^{01}_{\Sigma^{\prime\prime}}(y)&= -0.0397887e^{-2y}& W^{01}_{\Sigma^\prime}(y)&= -0.0795775 e^{-2y} &\nonumber\\
\end{flalign}

\subsection*{Carbon ($^{12}$C)}
\begin{flalign}
W^{00}_{M}(y)&= 0.565882 e^{-2y} (2.25 - y)^2& \nonumber\\ 
W^{00}_{\Phi^{\prime\prime}}(y)&= 0.0480805 e^{-2y} & \nonumber\\
W^{00}_{M\Phi^{\prime\prime}}(y)&= e^{-2y} (-0.371134 + 0.164948 y) & 
\end{flalign}

\subsection*{Fluorine ($^{19}$F)}
\begin{flalign}
W^{00}_{M}(y) &= e^{-2 y} \left(0.0662231 y^4-1.23196 y^3+7.68018 y^2-18.1437 y+14.3637\right)&\nonumber\\ 
W^{11}_{M}(y) &= e^{-2 y} \left(0.00683961 y^4-0.043991 y^3+0.103729 y^2-0.106103 y+0.0397887\right)&\nonumber\\ 
W^{10}_{M}(y) &= e^{-2 y} \left(-0.0212824 y^4+0.266402 y^3-1.00139 y^2+1.48545 y-0.755986\right)&\nonumber\\ 
W^{01}_{M}(y) &= e^{-2 y} \left(-0.0212824 y^4+0.266402 y^3-1.00139 y^2+1.48545 y-0.755986\right)&\nonumber\\ 
W^{00}_{\Sigma''}(y) &=e^{-2 y} \left(0.00679021 y^4-0.0400138 y^3+0.0896114 y^2-0.0903448 y+0.0346155\right)&\nonumber\\ 
W^{11}_{\Sigma''}(y) &=e^{-2 y} \left(0.00716698 y^4-0.0431855 y^3+0.0977332 y^2-0.0984534 y+0.0372496\right)&\nonumber\\ 
W^{10}_{\Sigma''}(y) &=e^{-2 y} \left(0.00697605 y^4-0.041572 y^3+0.0935814 y^2-0.0943139 y+0.0359084\right)&\nonumber\\ 
W^{01}_{\Sigma''}(y) &=e^{-2 y} \left(0.00697605 y^4-0.041572 y^3+0.0935814 y^2-0.0943139 y+0.0359084\right)&\nonumber\\ 
W^{00}_{\Sigma'}(y) &=e^{-2 y} \left(0.0149629 y^4-0.0867403 y^3+0.19008 y^2-0.186579 y+0.069231\right)&\nonumber\\ 
W^{11}_{\Sigma'}(y) &=e^{-2 y} \left(0.0140042 y^4-0.0865149 y^3+0.198218 y^2-0.199543 y+0.0744992\right)&\nonumber\\ 
W^{10}_{\Sigma'}(y) &=e^{-2 y} \left(0.0144756 y^4-0.0866714 y^3+0.194127 y^2-0.192953 y+0.0718168\right)&\nonumber\\ 
W^{01}_{\Sigma'}(y) &=e^{-2 y} \left(0.0144756 y^4-0.0866714 y^3+0.194127 y^2-0.192953 y+0.0718168\right)&\nonumber\\ 
W^{00}_{\Phi''}(y) &=e^{-2 y} \left(0.00314736 y^2-0.0157368 y+0.019671\right)&\nonumber\\ 
W^{11}_{\Phi''}(y) &=e^{-2 y} \left(0.000600385 y^2-0.00300193 y+0.00375241\right)&\nonumber\\ 
W^{10}_{\Phi''}(y) &=e^{-2 y} \left(-0.00137464 y^2+0.00687319 y-0.00859149\right)&\nonumber\\ 
W^{01}_{\Phi''}(y) &=e^{-2 y} \left(-0.00137464 y^2+0.00687319 y-0.00859149\right)&\nonumber\\ 
W^{00}_{\Delta}(y) &=3.60124\times10^{-6} e^{-2 y} \left(2.5 - y\right)^2&\nonumber\\ 
W^{11}_{\Delta}(y) &=0.000546075 e^{-2 y} (2.5 - y)^2&\nonumber\\ 
W^{10}_{\Delta}(y) &=0.0000443458 e^{-2 y} (2.5 - y)^2&\nonumber\\ 
W^{01}_{\Delta}(y) &=0.0000443458 e^{-2 y} (2.5 - y)^2&\nonumber\\ 
W^{00}_{M\Phi^{\prime\prime}}(y)&=e^{-2 y} \left(0.014437 y^3-0.17038 y^2+0.54834 y-0.531554\right)&\nonumber\\ 
W^{11}_{M\Phi^{\prime\prime}}(y)&=e^{-2 y} \left(0.00202642 y^3-0.0115829 y^2+0.0211796 y-0.012219\right)&\nonumber\\ 
W^{10}_{M\Phi^{\prime\prime}}(y)&=e^{-2 y} \left(-0.00463969 y^3+0.02652 y^2-0.0484926 y+0.0279765\right)&\nonumber\\ 
W^{01}_{M\Phi^{\prime\prime}}(y)&=e^{-2 y} \left(-0.0063055 y^3+0.0744149 y^2-0.239492 y+0.232161\right)&
\end{flalign}

\subsection*{Sulfur ($^{32}$S)} 
\begin{flalign}
W^{00}_{M}(y)&=  0.580305 e^{-2y} (5.92494 - 5.43118 y + y^2)^2&\nonumber\\ 
W^{00}_{\Phi^{\prime\prime}}(y)&= 0.0765941 e^{-2y} (2.5 - y)^2&\nonumber\\
W^{00}_{M\Phi^{\prime\prime}}(y)&= e^{-2y} (-3.12284 + 4.11173 y - 1.6721 y^2 + 0.210827 y^3)&
\end{flalign}

%\bibliography{ref}{}
%\bibliographystyle{jhep.bst}

\providecommand{\href}[2]{#2}\begingroup\raggedright\endgroup

\end{document}